\documentclass[pdftex, twocolumn, epjc3]{svjour3}
\pdfoutput=1

\usepackage{placeins}
\usepackage{multirow}
\usepackage[utf8]{inputenc}
\usepackage{color}

\RequirePackage[detect-all=true,group-digits=true,group-separator={,},binary-units=true]{siunitx}
\RequirePackage{xspace}
\RequirePackage{graphicx}
\usepackage[pdftex,bookmarks,hidelinks]{hyperref}
\usepackage{lineno}

\def\bracketbar{\hbox{\kern-9pt\raise1pt%
    \hbox{{\tiny(}{\lower1.5pt\hbox{\bf--}}{\tiny)}}}}

\journalname{Eur. Phys. J. C}
\hypersetup{draft}

\newcommand{\snowglobes}{{\sf{SNOwGLoBES}}}
\begin{document}
\title{Supernova Neutrino Burst Detection with the Deep Underground Neutrino Experiment
}

\date{\today}

\authorrunning{DUNE Collaboration}
\author{B.~Abi\thanksref{Oxford}
	 \and R.~Acciarri\thanksref{Fermi}
	 \and M.~A.~Acero\thanksref{Atlantico}
	 \and G.~Adamov\thanksref{Georgian}
	 \and D.~Adams\thanksref{Brookhaven}
	 \and M.~Adinolfi\thanksref{Bristol}
	 \and Z.~Ahmad\thanksref{VariableEnergy}
	 \and J.~Ahmed\thanksref{Warwick}
	 \and T.~Alion\thanksref{Sussex}
	 \and S.~Alonso Monsalve\thanksref{CERN}
% ...
}
\institute {
% ...
\pagebreak[0] Universidad del Atl{\'a}ntico, Atl{\'a}ntico, Colombia\label{Atlantico}
% ...
\and\pagebreak[0] University of Bristol, Bristol BS8 1TL, United Kingdom\label{Bristol}
	 \and\pagebreak[0] Brookhaven National Laboratory, Upton, NY 11973, USA\label{Brookhaven}
% ...
\and\pagebreak[0] CERN, The European Organization for Nuclear Research, 1211 Meyrin, Switzerland\label{CERN}
% ...
\and\pagebreak[0] Fermi National Accelerator Laboratory, Batavia, IL 60510, USA\label{Fermi}
% ...
\and\pagebreak[0] Georgian Technical University, Tbilisi, Georgia\label{Georgian}
% ...
\and\pagebreak[0] University of Oxford, Oxford, OX1 3RH, United Kingdom\label{Oxford}
% ...
\and\pagebreak[0] University of Sussex, Brighton, BN1 9RH, United Kingdom\label{Sussex}
% ...
\and\pagebreak[0] Variable Energy Cyclotron Centre, 700 064 West Bengal, India\label{VariableEnergy}
% ...
\and\pagebreak[0] University of Warwick, Coventry CV4 7AL, United Kingdom\label{Warwick}
% ...
}

\onecolumn
\maketitle
\twocolumn
\sloppy

\begin{abstract}
The Deep Underground Neutrino Experiment (DUNE), a 40-kton underground
liquid argon time projection chamber experiment, will be sensitive to
the electron-neutrino flavor component of the burst of neutrinos
expected from the next Galactic core-collapse supernova.  Such an
observation will bring unique insight into the astrophysics of core
collapse as well as into the properties of neutrinos.  The general
capabilities of DUNE for neutrino detection in the relevant few- to
few-tens-of-MeV neutrino energy range will be described.  As an
example, DUNE's ability to constrain the  $\nu_e$
spectral parameters of the neutrino burst will be considered.

\end{abstract}

\section{Introduction}

The Deep Underground Neutrino Experiment (DUNE) will be made up of
four 10-kton liquid
argon time projection chambers underground in South Dakota as part of
the DUNE/Long-Baseline Neutrino Facility (LNBF) program.   DUNE will
record and reconstruct
neutrino interactions in the $\sim$GeV and higher range for studies of
neutrino oscillation parameters and searches for new physics using
neutrinos from a 
beam sent from Fermilab and using neutrinos from the atmosphere.   DUNE's dynamic
range is such that  it is also sensitive to neutrinos with energies
down to about 5 MeV.  Charged-current (CC) interactions of neutrinos from around 5~MeV to several tens of MeV create short electron tracks in liquid argon, potentially accompanied by
gamma-ray and other secondary particle signatures.   
This regime is of
particular interest for detection of the burst of neutrinos from a galactic
core-collapse supernova.  Such a detection would be of great interest in the context of multi-messenger astronomy.  
The sensitivity of DUNE is primarily to electron-flavor
  neutrinos  from supernovae, and this capability is unique among existing and proposed supernova neutrino detectors for the next decades.  
Neutrinos and antineutrinos from other astrophysical sources, such as solar and diffuse  supernova background neutrinos, are also potentially detectable.  
This low-energy (few to few tens of MeV) event regime has particular reconstruction, background and triggering challenges.

One of the primary physics goals of DUNE as stated in the Technical
Design Report (TDR)~\cite{Abi:2020wmh,Abi:2020evt,Abi:2020loh} is to
``Detect and measure the $\nu_{e}$ flux from a core-collapse
      supernova within our galaxy, should one occur during the lifetime
      of the DUNE experiment. Such a measurement would provide a wealth
      of unique information about the early stages of core collapse, and
      could even signal the birth of a black hole."~\cite{Abi:2018dnh}.

This paper will document selected studies from the DUNE 
TDR aimed at
understanding DUNE's sensitivity to low-energy neutrino physics, with an
emphasis on supernova burst signals. Section~\ref{sec:snb-lowe-snb}
describes basic supernova neutrino physics, as well as prospects for astrophysics and particle physics from observation of a burst.  Section~\ref{sec:landscape} gives
an overview of the landscape of supernova neutrino burst detection.   Section~\ref{sec:dunedet}
gives a brief description of the DUNE far detector.  
Section~\ref{sec:lowe-events} describes the general properties of
low-energy events in DUNE, including interaction channels, simulation and reconstruction tools, and backgrounds.  The tools include MARLEY, a neutrino event generator
specifically developed for this energy regime~\cite{marley}, and the \snowglobes~ fast event-rate calculation tool~\cite{snowglobes}.  These are both open-source community tools, rather than DUNE-specific software.  The studies described here make use of MARLEY and \snowglobes~with input from the full DUNE simulation-reconstruction chain. Section~\ref{sec:sn-signals} describes the
expected supernova signal in DUNE, and Sec.~\ref{sec:burst_triggering} describes burst triggering studies (as distinct from offline reconstruction studies.)
Section~\ref{sec:flux-params} describes an example of a study of supernova flux parameter sensitivity in DUNE.
Details on
supernova pointing capabilities and solar neutrino capabilities will be described in  separate publications.

\section{Supernova neutrino bursts}
\label{sec:snb-lowe-snb}

The burst of neutrinos from the celebrated core-collapse supernova 1987A in the Large Magellanic Cloud, about
50~kpc from Earth, heralded the era of extragalactic neutrino
astronomy.  This single neutrino-based observation of a core collapse confirmed our basic understanding of its physical mechanism. Theoretical understanding of the process and of the potential to gain far deeper knowledge from a future observation has advanced considerably in the past decades.

\subsection{Neutrinos from Collapsed Stellar Cores: Basics}

A core-collapse supernova\footnote{``Supernova'' always
  refers to a ``core-collapse supernova'' in this paper, although we are aware that not all core collapses produce electromagnetically visible supernovae, and not all supernovae result from stellar core collapse.} occurs when a massive star reaches the end of its
life. As a result of nuclear burning throughout the star's life, the
central region of such a star gains an ``onion'' structure, with an
iron core at the center surrounded by concentric shells of lighter
elements (silicon, oxygen, neon, magnesium, carbon, etc). At
temperatures of $T\sim 10^{10}$ K and densities of $\rho \sim 10^{10}$
g/cm$^{3}$, the Fe core continuously loses energy by neutrino emission
(through pair annihilation and plasmon
decay~\cite{Braaten:1993jw}). Since iron cannot be further burned, the
lost energy cannot be replenished throughout the volume and the core
continues to contract and heat up, while also growing in mass thanks
to the shell burning. Eventually, the critical mass of about $1.4
M_{\odot}$ of Fe is reached, at which point a stable configuration is
no longer possible. As electrons are absorbed by the protons in nuclei
and some iron is disintegrated by thermal photons, the degeneracy pressure support is suddenly removed and the core collapses essentially in free fall, reaching speeds of about a quarter of the speed of light\footnote{Other collapse mechanisms are possible: an ``electron-capture'' supernova does not reach the final burning phase before highly degenerate electrons break apart nuclei and trigger a collapse.}.

The collapse of the central region is suddenly halted after $\sim
10^{-2}$ seconds, as the density reaches nuclear (or super-nuclear)
values. The central core rebounds and an outward-moving shock wave is formed. The extreme physical conditions of this core, in particular the densities of order $10^{12}-10^{14}$ g/cm$^{3}$, create a medium that is opaque even for neutrinos. As a consequence, the core initially has a trapped lepton number. The gravitational energy of the collapse at this stage is stored mostly in the degenerate Fermi sea of electrons ($E_{F}\sim 200$ MeV) and electron neutrinos, which are in equilibrium with the former. The temperature of this core is not more than 30~MeV, which means the core is relatively cold. 

At the next stage, the trapped energy and lepton number both escape from the core, carried by the least interacting particles, which in the standard model are neutrinos.  Neutrinos and antineutrinos of all flavors are emitted in a time span of a few seconds (their diffusion time). The resulting central object then settles to a neutron star, or a black hole. A tremendous amount of energy, some~$10^{53}$ ergs, is released in $10^{58}$ neutrinos with energies $\sim 10$~MeV. A fraction of this energy is absorbed by beta reactions into the material behind the shock wave that then blasts away the rest of the star, creating, in many cases, a spectacular explosion.
Yet, from the energetics point of view, this visible explosion is but a tiny perturbation on the total event. Over 99\% of all gravitational binding energy of the $1.4 M_{\odot}$ collapsed core -- some 10\% of its rest mass -- is emitted in neutrinos. 

\subsection{Stages of the Explosion}

The core-collapse neutrino signal starts with a short, sharp
``neutronization'' (or ``break-out'') burst primarily composed of
$\nu_e$ from $e^- + p \rightarrow \nu_e + n$. These neutrinos are messengers of the shock front breaking through the neutrinosphere (the surface of neutrino trapping): when this happens, iron is disintegrated, the neutrino scattering rate drops and the lepton number trapped just below the original neutrinosphere is suddenly released. This quick and intense burst is followed by an
``accretion'' phase lasting some hundreds of milliseconds, depending on the progenitor star mass, as matter falls onto the collapsed core and the shock is stalled at the distance of  $\sim 200$ km. The gravitational binding energy of the accreting material is powering the neutrino luminosity during this stage. The later
``cooling'' phase over $\sim$10~seconds represents the main part of
the signal, over which the proto-neutron star sheds its trapped energy.  

The flavor content and spectra of the neutrinos emitted from the neutrinosphere change
throughout these phases, and the supernova's evolution can
be followed with the neutrino signal.

\begin{figure}[htbp]
\includegraphics[width=0.98\linewidth]{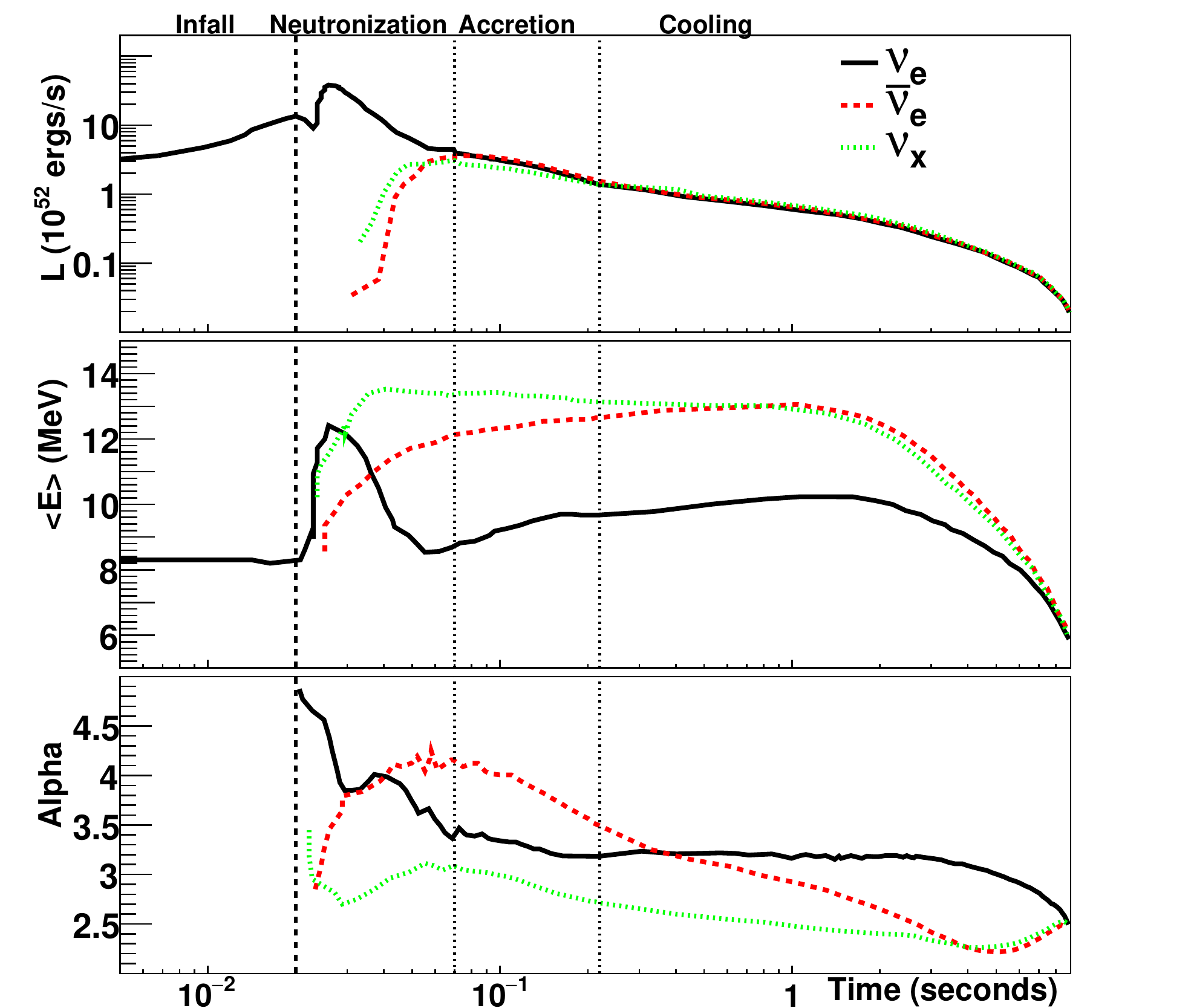}
\caption{Expected
  time-dependent flux parameters for a specific model for an
  electron-capture supernova~\cite{Huedepohl:2009wh}.  No
  flavor transitions are assumed. The
  top plot shows the luminosity as a function of time, the second plot
  shows average neutrino energy, and the third plot shows the $\alpha$
  (pinching) parameter.  The vertical dashed line at 0.02 seconds indicates
  the time of core bounce, and the vertical lines indicate different
  eras in the supernova evolution.  The leftmost time interval
  indicates the infall period.  The next interval, from core bounce to
  50~ms, is the neutronization burst era, in which the flux is
  composed primarily of $\nu_e$.  The next period, from 50 to 200~ms,
  is the accretion period. The final era, from 0.2 to 9~seconds, is
  the proto-neutron-star cooling period.  The general features are
  qualitatively similar for most core-collapse supernova models.\label{fig:params}}

\end{figure}

\begin{figure*}
\centerline{\includegraphics[width=10cm]{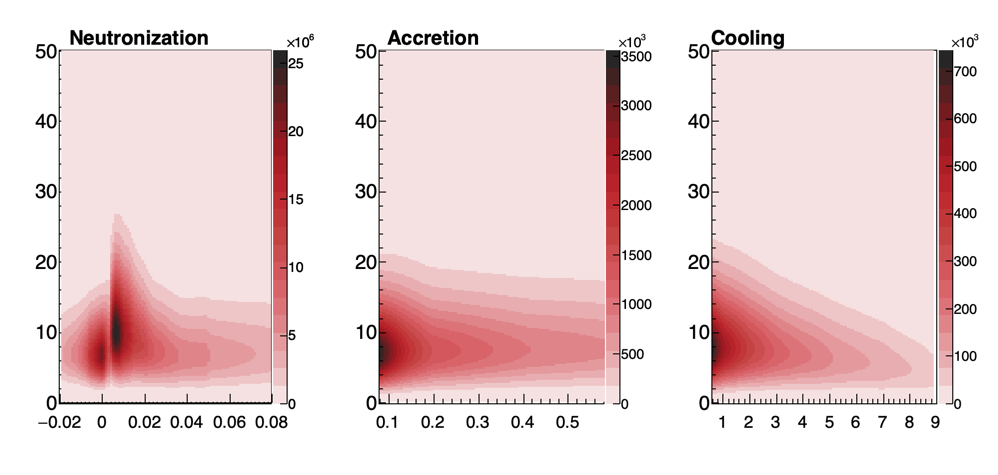}}
\centerline{\includegraphics[width=10cm]{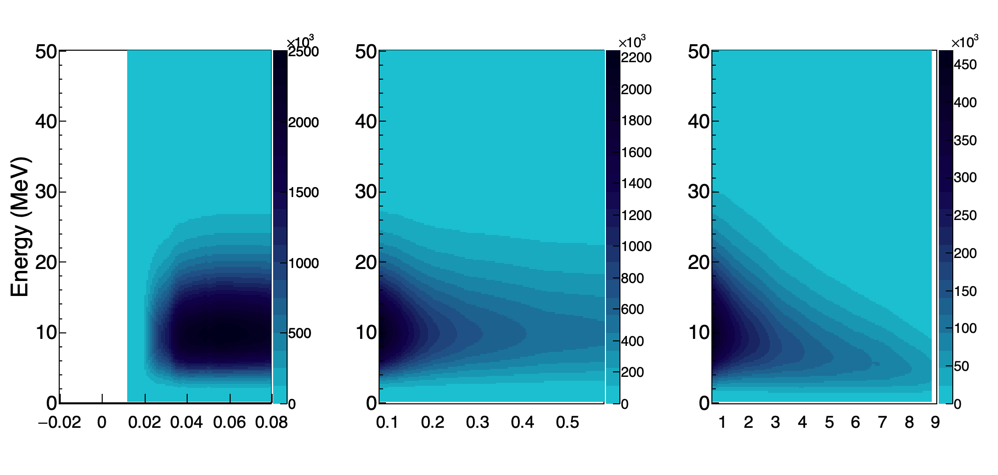}}
\centerline{\includegraphics[width=10cm]{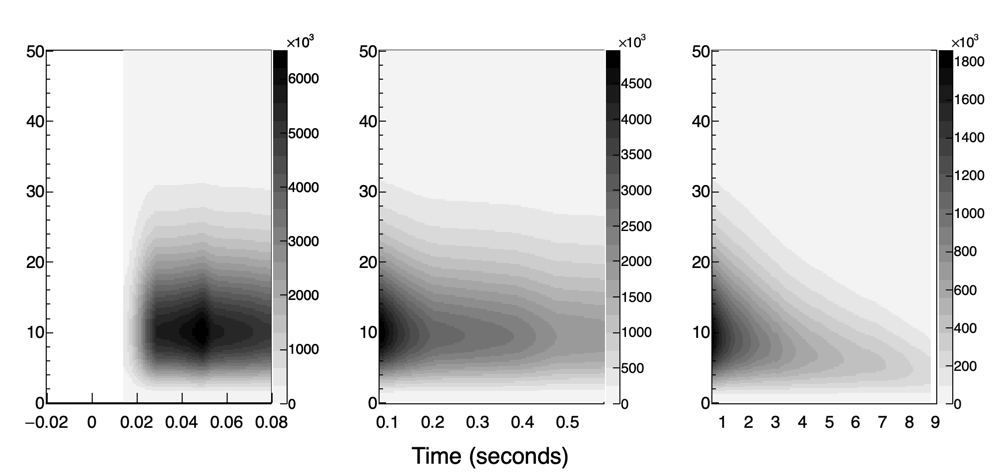}}
\caption{Examples of time-dependent neutrino spectra for the
    electron-capture supernova model~\cite{Huedepohl:2009wh}
    parameterized in Fig.~\ref{fig:params}, on three different
    timescales.  The x-axis for all plots indicates time in seconds and the y-axis indicates neutrino energy in MeV.    The z-axis color-shading units are neutrinos per cm$^2$ per
    millisecond per 0.2~MeV.  Note the different z scales in the panels. Core bounce is at $t=0$.  Top: $\nu_e$.  Center: $\bar{\nu}_e$.
    Bottom: $\nu_x$.  Flavor transition effects are not included here; note they can have dramatic effects on the spectra. Figure modified from Ref.~\cite{stateoftheart}.\label{fig:3timescales}}
\end{figure*}

The physics of neutrino decoupling and spectra formation is far from trivial, owing to the energy dependence of the cross sections and the roles played by both CC and neutral-current (NC) reactions.
Detailed transport calculations using methods such as MC or Boltzmann solvers have been employed. It has been observed that flux spectra coming out of such simulations can typically be parameterized at a given moment in time by the following ansatz (e.g.,~\cite{Minakata:2008nc,Tamborra:2012ac}):
\begin{equation}
        \label{eq:pinched}
        \phi(E_{\nu}) = \mathcal{N} 
        \left(\frac{E_{\nu}}{\langle E_{\nu} \rangle}\right)^{\alpha} \exp\left[-\left(\alpha + 1\right)\frac{E_{\nu}}{\langle E_{\nu} \rangle}\right] \ ,
\end{equation}
where $E_{\nu}$ is the neutrino energy, $\langle E_\nu \rangle$ is the
mean neutrino energy, $\alpha$ is a ``pinching parameter'', and
$\mathcal{N}$ is a normalization constant related to the total luminosity.
Large $\alpha$ corresponds to a more ``pinched'' spectrum (suppressed
tails at high and low energy). This parameterization is referred to as a
``pinched-thermal'' form. The different $\nu_e$, $\overline{\nu}_e$ and
$\nu_x~(x = \mu, \tau, \bar{\mu},\bar{\tau}$) flavors are expected to have different
average energy and $\alpha$ parameters and to evolve differently in
time. 

The initial spectra get further processed by flavor transitions, and understanding these oscillations is very important for extracting physics from the detected signal (see Sec.~\ref{sec:mh}).

In general, one can describe the neutrino flux as a function of time by specifying the three pinching parameters in successive time slices.
 Figure~\ref{fig:params} gives an example of pinching parameters as a function of time for a specific model, and Fig.~\ref{fig:3timescales} shows the spectra for the three flavors as a function of time corresponding to this parameterized description.  We have verified that the time-integrated spectrum for each flavor is expected to be reasonably well approximated by the pinched-thermal form as well.

\subsection{Astrophysical Observables}

A number of astrophysical phenomena associated with supernovae are expected to be observable
in the supernova neutrino signal, providing a remarkable window into the event.  In particular, the supernova explosion mechanism, which in the current paradigm involves energy deposition into the stellar envelope via neutrino interactions, is still not well understood, and the neutrinos themselves will bring the insight needed to confirm or refute the paradigm.

There are many other examples of astrophysical observables:
\begin{itemize}
\item The initial burst, primarily composed of $\nu_e$ and called the
  ``neutronization'' or ``breakout''
  burst, 
  represents only a small component of the total signal.  However,
  flavor transition effects can manifest themselves in an observable manner
  in this burst, and flavor transformations can be modified by the
  ``halo'' of neutrinos generated in the supernova envelope by
  scattering~\cite{Cherry:2013mv}.
\item The formation of a black hole would cause a sharp signal cutoff
  (e.g.,~\cite{Beacom:2000qy,Fischer:2008rh,Li:2020ujl}).
\item Shock wave effects (e.g.,~\cite{Schirato:2002tg}) would cause a
  time-dependent change in flavor and spectral composition as the
  shock wave propagates.
\item The standing accretion shock instability
  (SASI)~\cite{Hanke:2011jf,Hanke:2013ena}, a ``sloshing'' mode
  predicted by three-dimensional neutrino-hydrodynamics simulations of
  supernova cores, would give an oscillatory flavor-dependent
  modulation of the flux.
\item Turbulence effects~\cite{Friedland:2006ta,Lund:2013uta} would
  also cause flavor-dependent spectral modification as a function of
  time.

\end{itemize}

The supernova neutrino burst is prompt with respect to the
electromagnetic supernova signal and therefore can be exploited to provide an
early warning to astronomers~\cite{Antonioli:2004zb,Scholberg:2008fa}.  
Note that not every core collapse will produce an observable supernova, and observation of a neutrino burst in the absence of an electromagnetic event would be very interesting. 

Even non-observation of a burst, or non-observation of
a $\nu_e$ component of a burst in the presence of supernovae (or other
astrophysical events) observed in electromagnetic or gravitational
wave channels, would still provide valuable information about the
nature of the sources.  Furthermore, a long-timescale, sensitive search
yielding no bursts will also provide limits on the rate of
core-collapse supernovae.

Observation of a supernova neutrino burst in coincidence with gravitational waves (which would also be prompt, and could indeed provide a time reference for a time-of-flight analysis) would be especially interesting~\cite{Arnaud:2003zr,Ott:2012jq,Mueller:2012sv,Nishizawa:2014zna}.

The better one can understand the astrophysical nature of core-collapse supernovae, the easier it will be to extract information about particle physics.  

\subsection{Prospects for Neutrino Physics and Other Particle Physics}
\label{sec:physics-snblowe-neutrino-physics}

A core-collapse supernova is essentially a gravity-powered neutrino bomb: the energy of the collapse is initially stored in the Fermi seas of electrons and neutrinos and then gradually leaked out by neutrino diffusion. The key property of neutrinos that makes them play such a dominant role in the supernova dynamics is the feebleness of their interactions. It then follows that should there be new light ($< 100$ MeV) particles with even weaker interactions, these could alter the energy transport process and the resulting evolution of the nascent proto-neutron star. Moreover, additional interactions or properties of neutrinos could also be manifested in this way. 

Thus, a core-collapse supernova can  be thought of as an extremely hermetic system, which can be used to search for numerous types of new physics (e.g.,~\cite{Schramm:1990pf,Raffelt:1999tx}). The list includes various Goldstone bosons (e.g., Majorons), neutrino magnetic moments, new gauge bosons (``dark photons''), ``unparticles'', and extra-dimensional gauge bosons. The existing data from SN1987A already provide significant constraints on these scenarios by confirming the basic energy balance of the explosion. At the same time, more precision is highly desirable and will be provided with the next galactic supernova. 

Such energy-loss-based analysis will make use of two types of information. First, the total energy of the emitted neutrinos should be compared with the expected release in the gravitational collapse.  Note that measurements of all flavors, including $\nu_e$, are needed for the best estimate of the energy release.
Second, the rate of cooling of the protoneutron state should be measured and compared with what is expected from diffusion of the standard neutrinos.
The detection of a supernova neutrino burst also allows
the exploration of corrections to the neutrino velocity that could arise
due to violations of Lorentz invariance \cite{km2012}.


The flavor transition physics and its signatures are a major part of
the physics program. Compared to the well-understood case of solar
neutrinos, in a supernova, neutrino flavor transformations are much
more involved. For supernovae, there are both neutrinos and antineutrinos, and the density profile is such that
both mass splittings---``solar" and ``atmospheric" --- have an effect on the neutrino propagation.
While flavor transitions can be reasonably well understood during 
early periods of the neutrino emission as standard
Mikheyev-Smirnov-Wolfenstein (MSW)
transitions in the varying density profile of the overlying material, during
later periods the physics of the transformations is significantly richer.
For example, several seconds after the onset of the explosion, the
flavor conversion probability is affected by the expanding shock front
and the turbulent region behind it. The conversion process in such a
stochastic profile is qualitatively different from the adiabatic MSW
effect in the smooth, fixed density profile of the Sun~\cite{Mirizzi:2015eza}.

Even more complexity is brought about by the coherent scattering of neutrinos off each other. This neutrino ``self-refraction'' 
 results in highly nontrivial flavor transformations close to the neutrinosphere, typically within a few hundred kilometers from the center, where the density of streaming neutrinos is very high. Since the evolving flavor composition of the neutrino flux feeds back into the oscillation Hamiltonian, the problem is nonlinear. Furthermore, as the interactions couple neutrinos and antineutrinos of different flavors and energies, the oscillations are characterized by ``collective" modes.    This complexity leads to very rich physics that has been the subject of intense interest over the last decade and a voluminous literature exists exploring these collective phenomena,
e.g.,~\cite{Duan:2005cp,Fogli:2007bk,Raffelt:2007cb,Raffelt:2007xt,EstebanPretel:2008ni,Duan:2009cd,Dasgupta:2009mg,Duan:2010bg,Duan:2010bf,Wu:2014kaa}.  This is an active theoretical field and the effects are not yet fully understood. A supernova burst is the only opportunity to study neutrino-neutrino interactions experimentally.

Active-sterile neutrino transitions may also have observable effects~\cite{Peres:2000ic,Esmaili:2014gya,Tang:2020pkp}.

The new effects can imprint information about the inner
workings of the explosion on the signal. The flavor transitions can modulate
the characteristics of the signal (both event rates and spectra as a
function of time).
In particular, the flavor transitions can imprint distinctive non-thermal features on the energy spectra, potentially making it possible to disentangle the effects of flavor transformations and the physics of neutrino spectra formation. This in turn should help us learn about the development of the explosion during the crucial first 10 seconds.

\subsubsection{Mass Ordering}\label{sec:mh}

The neutrino mass
ordering affects the specific flavor composition in multiple ways
during the different eras of neutrino emission.  
References~\cite{Mirizzi:2015eza,Scholberg:2017czd} survey in some detail the
multiple signatures of mass ordering that will imprint themselves on
the flux.  For many of these, the $\nu_e$ component of the signal will
be critical to measure.    Some signatures of mass ordering are more robust than
others, in the sense that the assumptions are less subject to
theoretical uncertainties.  One of the more robust of these is the
early-time signal, including the neutronization burst.   At
early times, the matter potential is dominant over the
neutrino-neutrino potential, which means that standard MSW effects are
in play. 
The early neutronization-burst period is expected to be
dominated by adiabatic MSW transitions driven by the ``H-resonance''
for $\Delta m^2_{3\ell}$, the larger squared mass difference between mass states, for which the following
neutrino-energy-independent relations apply:

\begin{eqnarray}  
 F_{\nu_e} &=& F^0_{\nu_x} \,\ \,\ \,\ \,\ \,\ \,\ \,\ \,\  \,\ \,\ \,\ \,\   \,\ \,\  \,\ \,\ \,\ \,\ \,\ \,\  \,\ \,\ \textrm{(NO)} \,\ , \label{eq:msw_nmo}\\
 F_{\nu_e} &=&  \sin^2 \theta_{12} F^0_{\nu_e} +
\cos^2 \theta_{12} F^0_{\nu_x}  \,\ \,\ \,\ \,\ \textrm{(IO)} \,\,
\label{eq:msw_imo}
\end{eqnarray} 
 and 
\begin{eqnarray}  
 F_{\bar\nu_e} &=& \cos^2 \theta_{12} F^0_{\bar\nu_e} + \sin^2 \theta_{12} F^0_{\bar\nu_x}   \,\   \,\ \,\  \,\ \,\ \,\ \,\ \,\ \,\  \,\ \,\ \textrm{(NO)} \,\ , \label{eq:msw_nmo_anti}\\
 F_{\bar\nu_e} &=&   F^0_{\bar\nu_x}  \,\ \,\ \,\ \,\ \,\ \,\ \,\ \,\ \,\ \,\ \,\ \,\ \,\ \,\ \,\ \,\ 
 \,\ \,\ \,\ \,\ \,\ \,\ \,\ \,\ \,\ \,\ \,\ \,\ \,\  \,\
\textrm{(IO)} \,\,\label{eq:msw_imo_anti}
\end{eqnarray} 

where $F$s are the fluxes corresponding to the respective flavors,
and the $^0$ superscript represents flux before transition.
In this case, for the normal ordering (NO), the neutronization burst, which is
emitted as nearly pure $\nu_e$, is strongly suppressed, whereas for
the inverted ordering (IO), the neutronization burst is only partly suppressed.   An example of this effect is considered in Sec.~\ref{sec:event_rates} for DUNE's expected signal.

Of course, if the mass ordering is already known, one can turn the question around and use the terrestrial determination to better disentangle the other  particle physics and astrophysics knowledge from the observed signal.  A detailed investigation of mass-ordering effects over a range of models will be the topic of a future publication.

\section{The Supernova Burst Neutrino Detection Landscape}\label{sec:landscape}
.  
The few dozen recorded $\bar{\nu}_e$ events from SN1987A~\cite{Bionta:1987qt,Hirata:1987hu,Alekseev:1987ej}
have confirmed the basic physical
picture of core collapse and yielded constraints on a wide range of new
physics~\cite{Schramm:1990pf,Vissani:2014doa}.  The community anticipates much
more bountiful data and corresponding advances in knowledge when the next nearby star collapses.

Core-collapse supernovae within a few hundred kiloparsecs of Earth---
within our own Galaxy and nearby--- are quite rare on a human
timescale.  They are expected once every few decades in the Milky Way
(within about 20~kpc), and with a similar rate in Andromeda (about
780~kpc away.)  However, core collapses should be common enough to have
a reasonable chance of occurring during the few-decades-long lifetime
of a typical large-scale neutrino detector.  The rarity of these
spectacular events makes it all the more critical for the scientific community to
be prepared to capture every last bit of information from them.

In principle, the information in a supernova neutrino burst available 
to neutrino experimentalists is comprised of the flavor, energy and time structure of the several-tens-of-seconds-long, all-flavor,
few-tens-of-MeV neutrino burst~\cite{Mirizzi:2015eza,Horiuchi:2017sku}.  Imprinted on the neutrino spectrum as a function
of time is information about the progenitor, the collapse, the
explosion, and the remnant, as well as information about neutrino
parameters and potentially exotic new physics.  The neutrino energies
and flavor content of the burst can be measured only imperfectly due
to both the intrinsic nature of the weak interactions of neutrinos with matter
and to the imperfect detection resolution of any real detector.
For example, supernova burst energies are below CC
threshold for $\nu_\mu$, $\nu_\tau$, $\bar{\nu}_\mu$ and
$\bar{\nu}_{\tau}$ (collectively $\nu_x$), which represent two-thirds
of the flux; therefore these flavors are accessible only via NC interactions, which tend to have low cross sections and
indistinct detector signatures. These issues make a comprehensive
unfolding of neutrino flavor, time and energy structure from the
observed interactions a challenging problem.

Much has occurred since 1987, both for experimental and theoretical
aspects of supernova neutrino detection.
There has been huge progress in the modeling of supernova explosions,
and there have been many new theoretical insights about
neutrino oscillation and exotic collective effects that may occur in
the supernova environment.    Experimentally,
worldwide detection capabilities have increased enormously, such that
there will be orders of magnitude more neutrino interactions from a core collapse at the center
of the Galaxy, about 8~kpc away.

\subsection{Current Experimental Landscape}


In the world's current supernova neutrino flavor sensitivity
portfolio~\cite{Scholberg:2012id,Mirizzi:2015eza}, the sensitivity is primarily to $\bar{\nu}_e$
flavor, via inverse beta decay on free protons in water and scintillator detectors worldwide (\cite{Ikeda:2007sa,Abe:2016waf,Abbasi:2011ss,Eguchi:2002dm,Agafonova:2014leu,Monzani:2006jg,NOvA:2020dll,Wei:2015qga}.)  There is only minor sensitivity to the $\nu_e$
component of the flux, via elastic scattering on electrons (ES) and subdominant interaction channels on nuclei, as well as in small detectors (HALO~\cite{Duba:2008zz} and MicroBooNE~\cite{Abratenko:2020hfy}.)  However $\nu_e$ statistics are small and and it can be difficult to disentangle the flavor content.
The $\nu_e$ component carries with it particularly interesting
information (e.g., neutronization burst neutrinos
are created primarily as $\nu_e$.)  

\subsection{Projected Landscape in the DUNE Era}
The next generation of supernova neutrino detectors will be dominated by Hyper-Kamiokande~\cite{Abe:2018uyc},
JUNO~\cite{An:2015jdp} and DUNE.  Hyper-K
and JUNO are sensitive primarily to $\bar{\nu}_e$, and Hyper-K in particular will have
potentially enormous statistics.  The next-generation long-string
water detectors, IceCube Gen-2~\cite{Aartsen:2014njl} and KM3NeT~\cite{Adrian-Martinez:2016fdl}, will bring improved burst timing.   New tens-of-ton scale
noble liquid detectors such as DARWIN~\cite{Aalbers:2016jon} will
bring new all-flavor 
sensitivity via NC coherent elastic neutrino-nucleus scattering.  To this landscape,
DUNE will bring unique $\nu_e$ sensitivity via $\nu_e$ charged-current
($\nu_e$CC) interactions on argon nuclei. It will offer
a new opportunity to measure the $\nu_e$ content of the burst with
high statistics and good event reconstruction.

The past decade has also brought rapid evolution of
multi-messenger astronomy.  With the advent of the detection
of gravitational
waves as well as high-energy extragalactic neutrino detection in
IceCube, a broad community of physicists and astronomers are now
collaborating to extract maximum information from observation in a
huge range of electromagnetic wavelengths, neutrinos, charged particles
and gravitational waves.  This collaboration resulted in the
spectacular multi-messenger observation of a kilonova~\cite{kilonova}.  The
next core-collapse supernova will be potentially an even more spectacular multi-messenger
observation.  Worldwide neutrino detectors are currently participants
in SNEWS, the SuperNova Early Warning System~\cite{snews}, which will be
upgraded to have enhanced capabilities over the next few
years~\cite{Kharusi:2020ovw}.  Information from DUNE will enhance the SNEWS
network's reach.

Neutrino pointing information is vital for prompt multi-messenger
capabilities.  Only some supernova neutrino detectors have the ability
to point back to the source of neutrinos.  Imaging water Cherenkov
detectors like Super-K can do well at this task via directional
reconstruction of neutrino-electron ES events. However, other detectors
lack pointing ability due to intrinsic quasi-isotropy of the neutrino
interactions, combined with lack of detector sensitivity to
final-state directionality.  Like Super-K, DUNE is capable of pointing
to the supernova via the good tracking ability of its time projection chamber (TPC.)

Supernova neutrino detection is more of a collaborative than a
competitive game.  The more information gathered by detectors
worldwide, the more extensive the knowledge to be gained; the whole is
more than the sum of the parts.  The flavor sensitivity of DUNE is
highly complementary to that of the other detectors and will bring
critical information for reconstruction of the entire burst's flavor and
spectral content as a function of time~\cite{Ankowski:2016lab}.

\subsection{Beyond Core Collapse}
While a core-collapse burst is a known source of 
low-energy ($<$100 MeV) neutrinos, there are other potential
interesting sources of neutrinos in this energy range.  Nearby Type Ia
~\cite{Wright:2016gar,Wright:2016xma} or pair instability supernova~\cite{Wright:2017zyq} events may create bursts
as well, although they are expected to be fainter in neutrinos than
core-collapse supernovae.  Mergers of binary neutron stars and of neutron stars and black holes
will be low-energy neutrino sources~\cite{Caballero:2009ww,Kyutoku:2017wnb}, although the rate of these close
enough to detect (i.e., within the Galaxy) will be small.  There are also interesting
steady-state sources of low-energy neutrinos~--- in particular, there
may still be useful oscillation and solar physics information to
extract via measurement of the solar neutrino flux. DUNE will have the
unique capability of measuring solar neutrino energies event by event
with $\nu_e$CC interactions with large statistics, in contrast to
other detectors, which primarily make use of recoil spectra~\cite{Capozzi:2018dat,Ioannisian:2017dkx}.  The technical
challenge for solar neutrinos is overcoming radiological and
cosmogenic backgrounds, although preliminary studies are promising.
The diffuse supernova background neutrinos~\cite{Moller:2018kpn} are another interesting target; these
have higher energy than solar neutrinos, but are very challenging due to very low
event rate.  There may also be surprises in store, both from burst
and steady-state signals, enabled by unique DUNE liquid argon tracking
technology.

\section{The DUNE Detector}\label{sec:dunedet}

The DUNE detector is part of the DUNE/Long Baseline Neutrino Facility
program, which comprises a GeV-scale, high-intensity neutrino beam
produced at Fermilab, a precision near detector at the Fermilab site,
and underground liquid argon time projection chambers (LArTPCs) 1300
km away.  The DUNE LArTPCs will be located at the Sanford Underground
Research Facility in South Dakota at a depth of 1.5~km.  Physics
goals in addition to supernova burst physics of the DUNE/LBNF program include:
measurement of neutrino oscillation in the long-baseline beam, study
of atmospheric neutrinos, searches for beyond-the-standard-model
physics, and searches for baryon number violation.

As described in the TDR, DUNE will have four modules of 70-kton liquid argon mass in total, of which
40 kton will be fiducial mass (10-kton fiducial mass per module).  Note
that in principle relevant active mass may exceed the nominal fiducial mass for
supernova neutrinos in a burst.  DUNE is prototyping two types of
LArTPCs.  Single-phase (SP)
LArTPC technology is designed to have horizontal drift of 3.5~m with
wrapped-wire readout including two induction and one charge collection anode planes.  Dual-phase (DP) LArTPC technology has vertical
drift over 12~meters.  At the liquid-gas interface at the top of a DP module, drifted ionization charge
is amplified and collected.

Liquid argon scintillates at 128~nm, and in both single-phase and
dual-phase technologies, wavelength-shifted photons will be collected by photodetectors
(PD), in addition to ionization charge.  For the single-phase design,
light-trapping devices called X-ARAPUCAs~\cite{Cancelo:2018dnf,Machado:2018rfb} will be mounted between wire layers.  These employ
dichroic filters and use silicon photomultipliers for photon sensing.
For the dual-phase design, cryogenic wavelength-shifter-coated
photomultiplier tubes will be deployed on the bottom of the detector.

Both detector designs should have roughly similar capabilities for
low-energy physics.  Most studies described here were done under the
SP design assumption; however the DP design should
provide similar results.

The  DUNE/LBNF experimental facility, detectors and overall physics program are described in detail in Ref.~\cite{Abi:2018dnh}.  More detail about the SP detector design can be found in Ref.~\cite{Abi:2018alz} and more detail about the DP detector design can be found in Ref.~\cite{Abi:2018rgm}.

\section{Low-Energy Events in DUNE}\label{sec:lowe-events}

\subsection{Detection Channels}

\begin{figure}[htbp]
\includegraphics[width=0.98\linewidth]{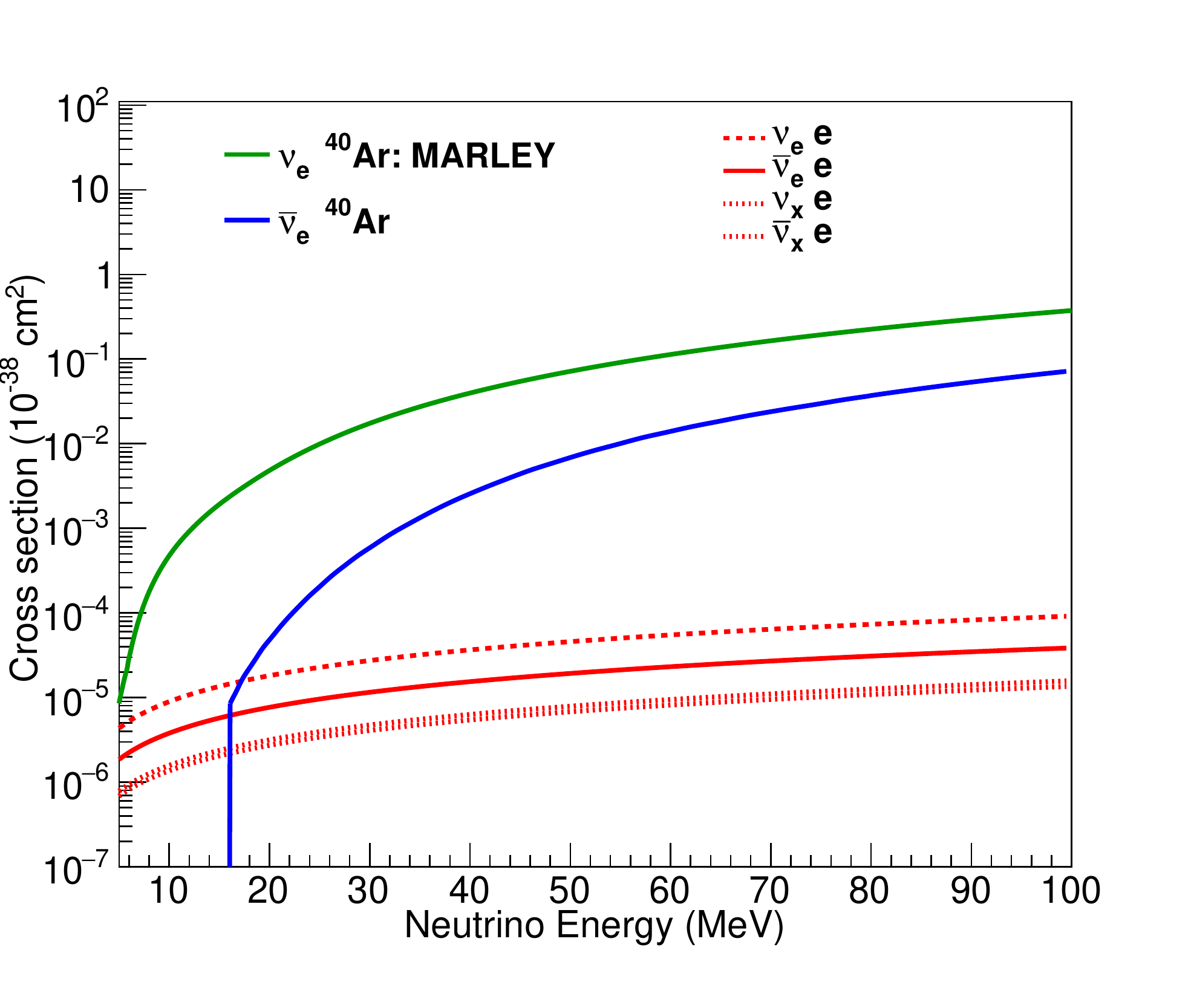}
\caption{Cross sections for supernova-relevant interactions in argon~\cite{snowglobes,GilBotella:2003sz} as a function of neutrino energy.  The $\nu_e$ CC cross section shown in green (used for the studies here) is from MARLEY (see Sec.~\ref{sec:marley}.)  Inelastic NC cross sections have large uncertainties and are not shown.\label{fig:xscns}}
\end{figure}

Liquid argon has a particular sensitivity to the $\nu_e$
component of a supernova neutrino burst, via the dominant interaction,
CC
absorption of $\nu_e$ on $^{40}$Ar,
\begin{equation}
\nu_e +{}^{40}{\rm Ar} \rightarrow e^- +{}^{40}{\rm K^*},
\label{eq:nueabs}
\end{equation}
for which the observable is the $e^-$ plus deexcitation products from
the excited $^{40}$K$^*$ final state.  Additional channels include a
$\bar{\nu}_e$ CC interaction and ES on electrons.
Cross sections for the most
relevant interactions are shown in Fig.~\ref{fig:xscns}.  It is worth
noting that none of the neutrino-$^{40}$Ar cross sections in this
energy range have been experimentally measured, although several
theoretical calculations exist~\cite{marley,marley2,snowglobes}.  The
uncertainties on the theoretical calculations are not generally
quantified, and they may be large.

Another process of interest for supernova detection in liquid argon detectors,
not yet fully studied,  is NC scattering on Ar nuclei by
any type of neutrino: $\nu_X +{\rm Ar} \rightarrow \nu_X +{\rm
  Ar}^*$,  for which the observable is the cascade of
deexcitation gammas from the final state Ar nucleus.  A dominant
9.8-MeV Ar$^*$ decay line has been recently identified as a spin-flip
M1 transition~\cite{Hayes}. At this energy the probability of
$e^+e^-$ pair production is relatively high, offering a potentially
interesting NC tag.  Other transitions are under
investigation.
NC interactions are not included in the studies presented here,
although they represent a topic of future investigation.

The predicted event rate from a supernova burst may be calculated by
folding expected neutrino flux differential energy spectra with cross
sections for the relevant channels, and with detector response; this
is done
using \snowglobes~\cite{snowglobes}~(see Sec.~\ref{snowglobes}.)

\subsection{Event Simulation and Reconstruction}

\begin{figure*}
\centering
\includegraphics[width=0.4\textwidth]{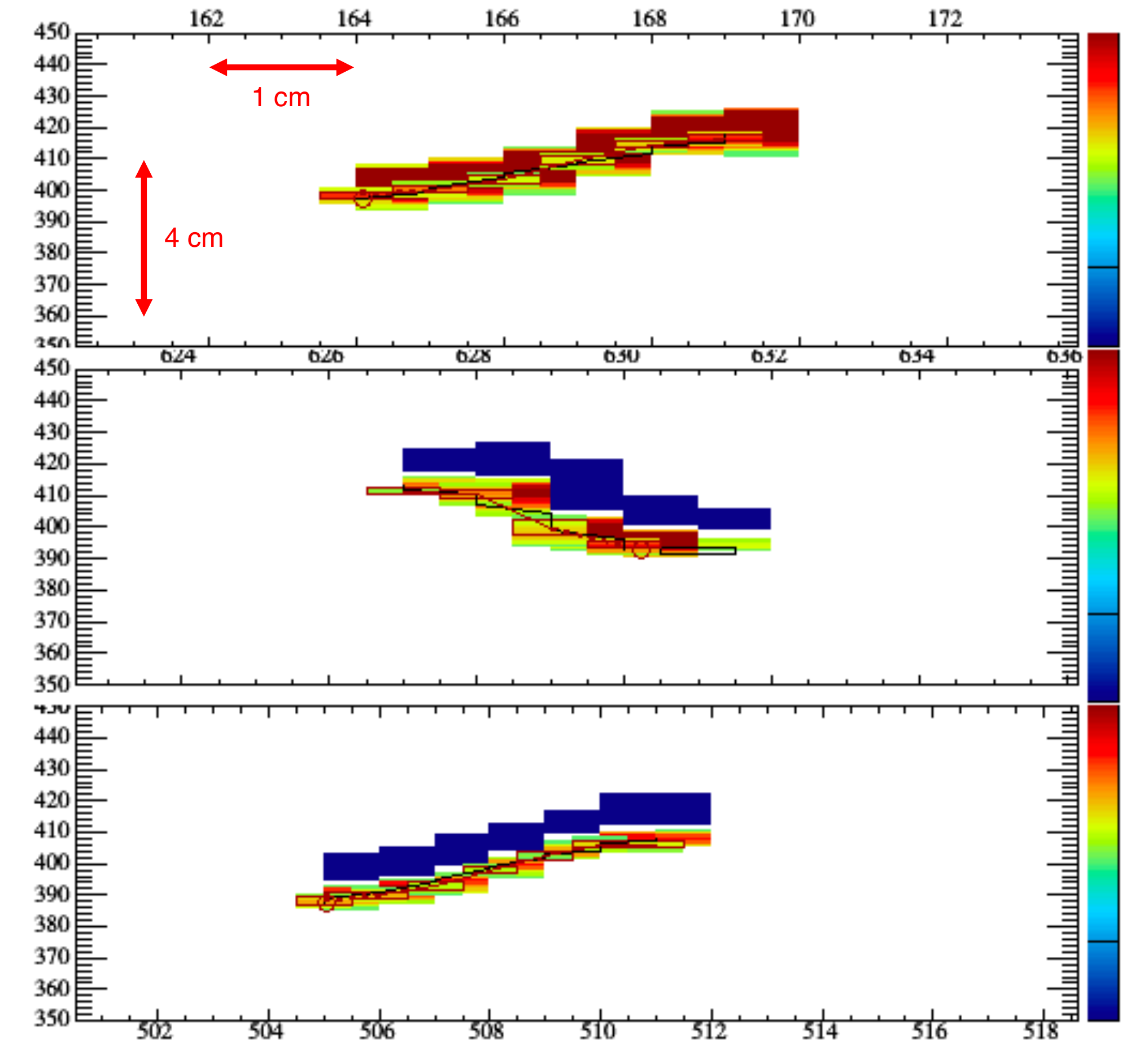}
\includegraphics[width=0.4\textwidth]{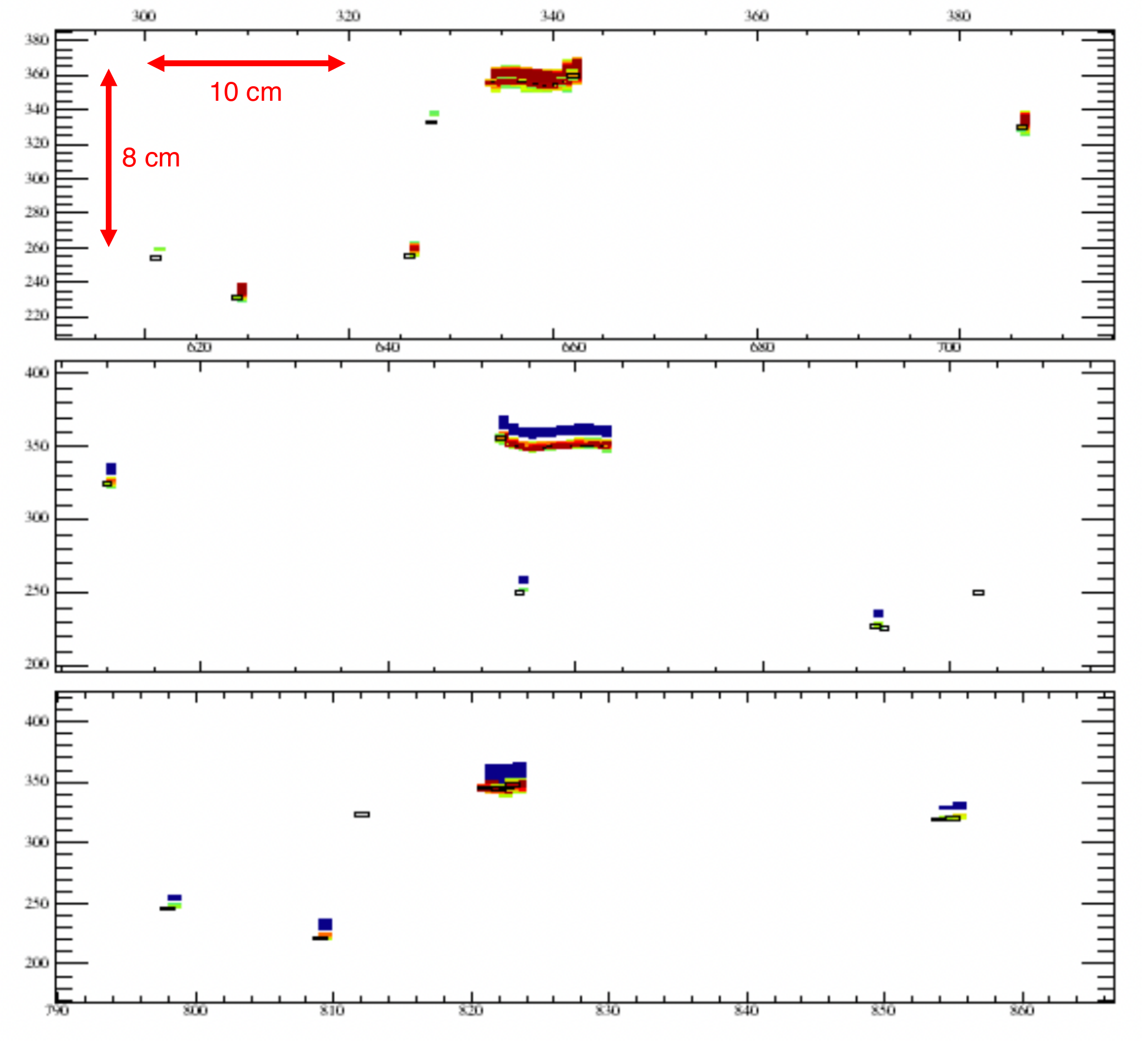}
\caption{Left: DUNE event display showing a simulated neutrino-electron ES event (10.25~MeV electron) with track reconstruction.  The vertical dimension indicates time and the horizontal dimension indicates wire number.  Color represents charge. The top panel shows the collection plane and the bottom panels show induction planes. The boxes represent reconstructed hits.  Right: simulated $\nu_e$CC event (20.25~MeV neutrino), showing electron track and blips from Compton-scattered gammas. The events have different spatial scales, as indicated on the figures. \label{fig:evdisplays} }
\end{figure*}

Supernova neutrino events, due to their low energies, will manifest
themselves primarily as spatially small events, perhaps up to a few tens of cm
scale, with stub-like tracks from
electrons (or positrons from the rarer $\bar{\nu}_e$ interactions).
Events from $\nu_e$CC, $\nu_e+{}^{40}{\rm
  Ar}\rightarrow e^{-}+{}^{40}{\rm K}^{*}$, are likely to be
accompanied by de-excitation products --- gamma rays and/or ejected
nucleons. Gamma rays are in principle observable via energy deposition
from Compton scattering, which will show up as small charge blips in
the time projection chamber.   Gamma rays can also be produced by
bremsstrahlung energy loss of electrons or positrons.  The critical
energy for bremsstrahlung energy loss for electrons in argon is about 45~MeV.
Ejected nucleons may result in loss of observed energy for
the event, although some may interact to produce observable deexcitations via
inelastic scatters on argon.  Such MeV-scale activity associated with neutrino interactions has been observed in the ArgoNeuT LArTPC~\cite{PhysRevD.99.012002}.
ES on electrons will result in single
scattered electron tracks, and single or cascades of gamma rays may result from NC 
excitations of the argon nucleus.   Each interaction category has, in principle, a
distinctive signature. 
Figure~\ref{fig:evdisplays} shows examples of simulated $\nu_e$CC and neutrino-electron ES interactions in DUNE.

The canonical event reconstruction task is to identify the interaction
channel, the neutrino flavor for CC events, and to determine the
four-momentum of the incoming neutrino; this overall task is the same for
low-energy events as for high-energy ones.  The challenge is to
reconstruct the properties of the lepton (if present), and to the extent
possible, to tag the interaction channel by the pattern of final-state
particles. LArSoft~\cite{larsoft.org} open-source event
simulation and
reconstruction software tools for low-energy events is employed; a full
description of the algorithms is beyond the scope of this work.  Performance is
described in Sec.~\ref{sec:performance}.  Enhanced tools are under
development, for example for interaction channel tagging; however,
standard tools already provide reasonable capability for energy
reconstruction and tracking of low-energy events.  Event
reconstruction in this energy range has been demonstrated by
MicroBooNE for Michel electrons~\cite{Acciarri:2017sjy}.

\subsubsection{Event Generation}\label{sec:marley}

MARLEY (Model of Argon Reaction Low Energy Yields)~\cite{marley,marley2} simulates tens-of-MeV
neutrino-nucleus interactions in liquid argon. For the studies here, MARLEY was only used to
simulate CC $\nu_e$ scattering on $^{40}$Ar, but other
reaction channels will be added in the future.

MARLEY weights the incident neutrino spectrum according to the assumed interaction cross section, selects an initial excited state
of the residual $^{40}$K$^*$ nucleus, and samples an outgoing electron
direction using the allowed approximation for the $\nu_e$CC differential cross
section, i.e., the zero momentum transfer and zero nucleon velocity
limit of the tree-level $\nu_e$CC differential cross section, which may be
written as

\begin{eqnarray*}
\frac{d\sigma}{d\cos \theta}
  & = \frac{G_F^2 |V_{ud}|^2}{2\pi} |\mathbf{p}_e|\, E_e \,F(Z_f, \beta_e) \\
 &\times \left[(1+\beta_e \cos\theta) B(F)   + \left(\frac{3 - \beta_e \cos\theta}
{3}\right)B(GT)\right].
\end{eqnarray*}

In this expression, $\theta$ is the angle between the incident neutrino and the
outgoing electron, $G_F$ is the Fermi constant, $V_{ud}$ is the quark mixing
matrix element, $F(Z_f, \beta_e)$ is the Fermi function, and $|\mathbf{p}_e|$,
$E_e$, and $\beta_e$ are the outgoing electron's three-momentum, total energy,
and velocity, respectively. $B(F)$ and $B(GT)$ are the Fermi and Gamow-Teller
matrix elements.
MARLEY computes this cross section using a table of Fermi and Gamow-Teller
nuclear matrix elements. Their values are taken from experimental measurements
at low excitation energies and a quasiparticle random phase approximation
(QRPA) calculation at high excitation energies. 

\begin{figure}[htbp]
\includegraphics[width=0.98\linewidth]{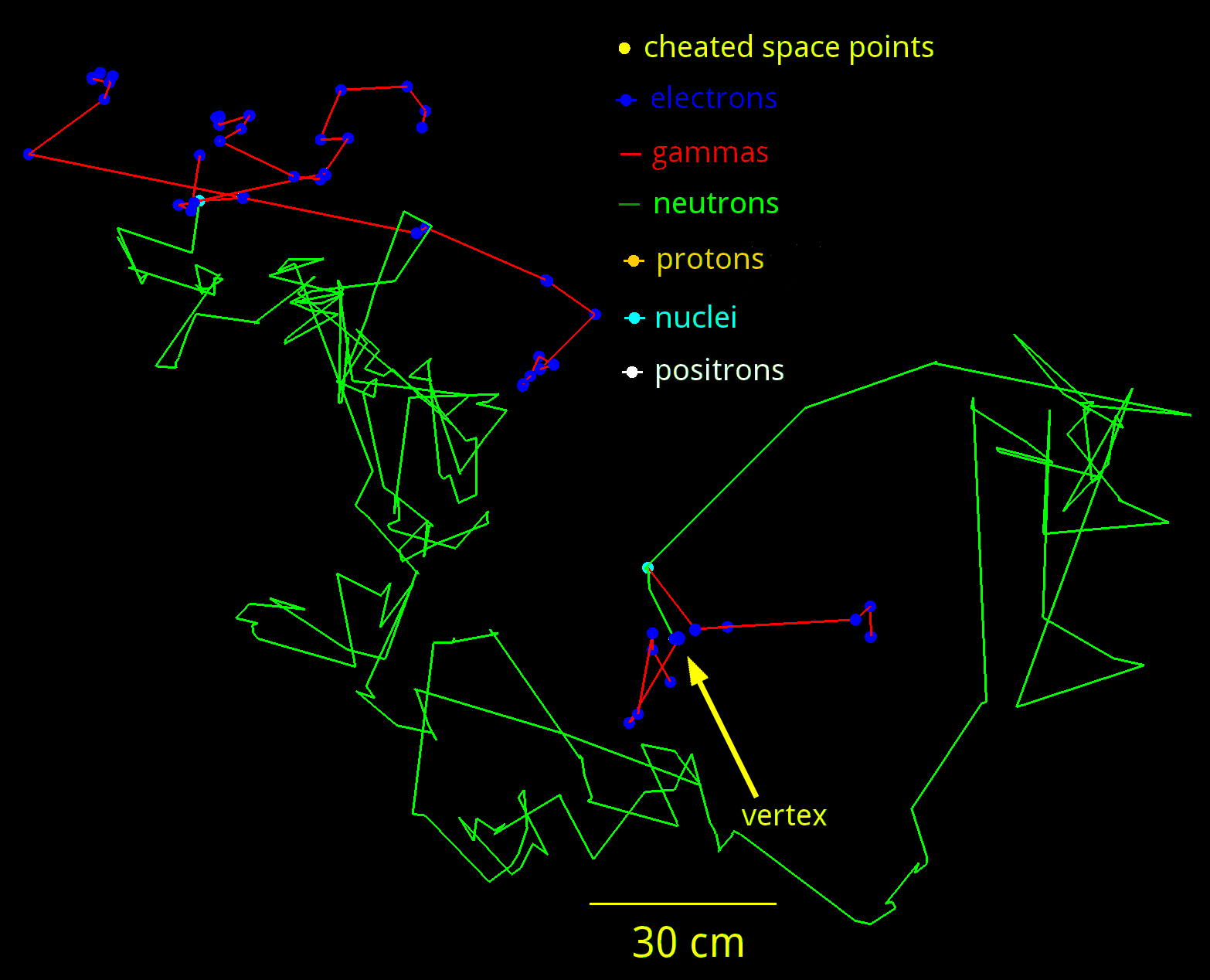}
\caption{Visualization of an
    example MARLEY $\nu_e$CC event simulated in LArSoft, showing the trajectories and energy
    deposition points of the interaction products. \label{fig:marleydist}}
\end{figure}

\begin{figure}[htbp]
\includegraphics[width=1.0\linewidth]{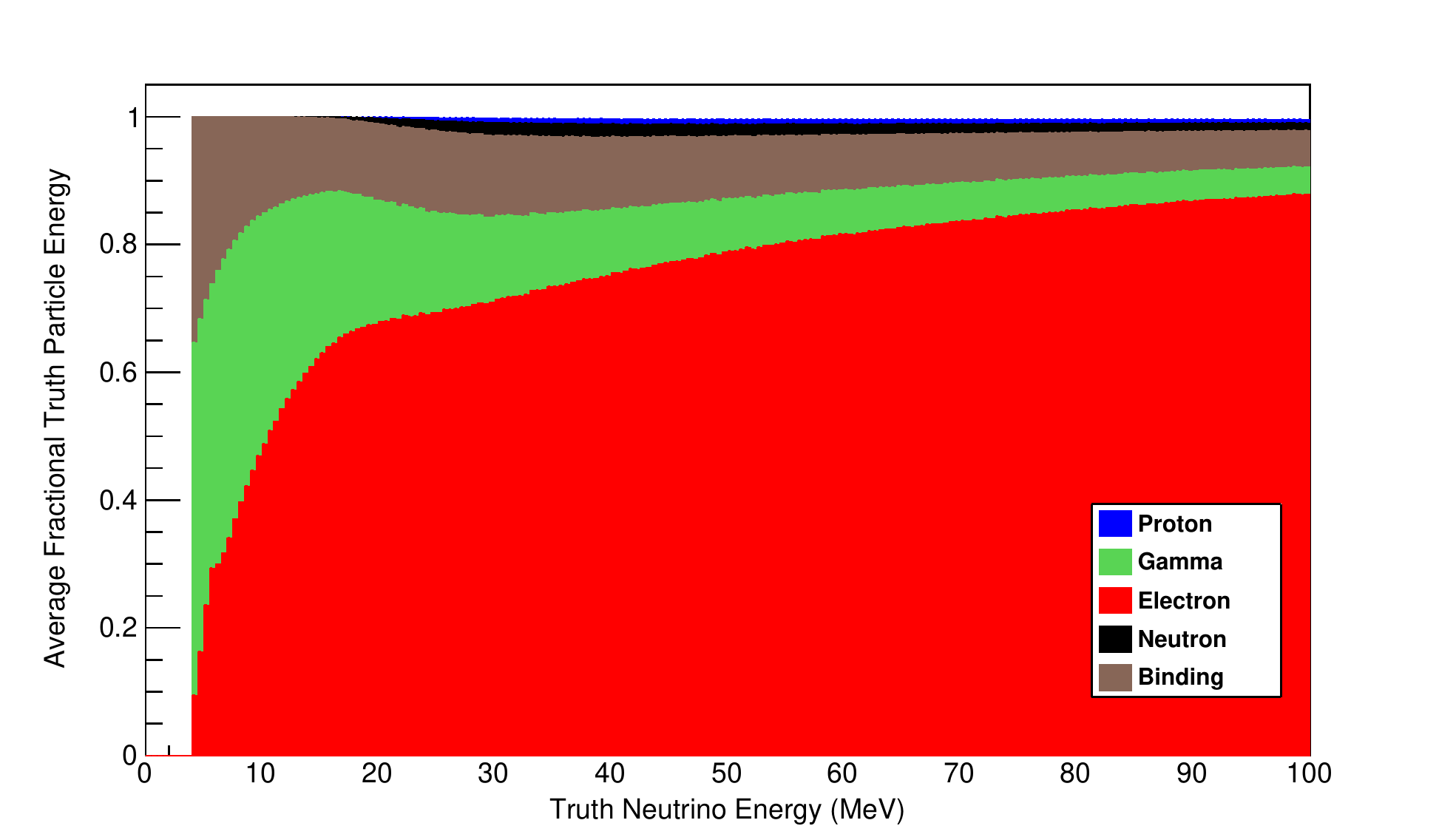}
\caption{ \label{fig:marleyfracenergy} Fraction of incident neutrino energy going to each final-state particle type in the MARLEY simulation as a function of neutrino energy. ``Binding energy" represents the difference in mass of the initial- and final-state nuclei, representing the kinematic threshold for the CC interaction. }
\end{figure}

After simulating the initial two-body $^{40}${Ar}($\nu_e$,
$e^{-}$)$^{40}$K$^*$ reaction for an event, MARLEY
also handles the subsequent nuclear de-excitation. For bound nuclear
states, the de-excitation gamma rays are sampled using tables of
experimental branching ratios~\cite{talys,Bhattacharya:1998hc,Cheoun:2012ha}. These tables are supplemented with
theoretical estimates when experimental data are unavailable. For
particle-unbound nuclear states, MARLEY simulates the competition between
gamma-ray and nuclear fragment\footnote{Nucleons and light nuclei up to
$^{4}${He} are considered.} emission using the Hauser-Feshbach
 statistical model.   Figure~\ref{fig:marleydist} shows an example
visualization of a simulated MARLEY event.  Figure~\ref{fig:marleyfracenergy} shows the mean fraction of energy apportioned to the different possible interaction products by MARLEY as a function of neutrino energy.

\subsubsection{Low-energy Event Reconstruction Performance}\label{sec:performance}

The LArSoft~\cite{larsoft.org} Geant4-based software package is used to simulate the
final-state products from MARLEY in the DUNE LArTPC.  Both TPC
ionization-based signals
and scintillation photon signals are simulated.

For the studies described  here, the DUNE LArSoft $1\times 2\times 6$~m far detector geometry was used~\cite{Abi:2020loh}, along with standard DUNE reconstruction tools included in the LArSoft package.  To determine event-by-event reconstruction information, 2D hits are formed using the HitFinder algorithm.  HitFinder scans through wires and defines hits in regions between two signal minima where the maximum signal is above threshold. The algorithm then performs $n$ Gaussian fits for $n$ consecutive regions.  The hit center is defined as the fitted Gaussian center, while the beginning and end are defined using the fitted Gaussian width.  We used the TrajCluster algorithm to form reconstructed clusters. The TrajCluster algorithm creates clusters using local information from 2D trajectories, taking advantage of minimal ionization energy loss compared to the kinetic energy of the particle. A 2D trajectory is formed from trajectory points defined by the cryostat, plane, and TPC in which the trajectory resides. The trajectory points are made up of charge-weighted positions of all hits used to form the point. The algorithm steps through the 2D space of hits sorted by wire ID number, region of interest in time, and then by ``multiplet” (i.e., a collection of hits found using a multi-Gaussian fit). Clusters are formed in the algorithm by stitching together nearby 2D hits. 3D track information is produced using the Projection Matching Algorithm (PMA). PMA takes in 2D clusters formed through TrajCluster, and the algorithm matches clusters in the three 2D projection wire planes to build the tracks. PMA measures the distance between projections, and tracks are formed based on stitching together nearby projections.

The photon (scintillation) simulation implemented ARAPUCA light collection devices with realistic light yields that differ between particle types.
Reconstructed photon flashes are used to correct 
ionization charge loss during drift, which provides substantial improvement to energy reconstruction.  Even in the absence of efficient TPC-flash matching, resolution smearing due to drift losses may end up being a small effect, particularly given the high electron lifetimes recently achieved in the DUNE prototype detector~\cite{Abi:2020mwi}.  Photons may also be used for calorimetry, although that method has not been implemented for these studies.

\begin{figure*}
\centering
\includegraphics[width=0.45\textwidth]{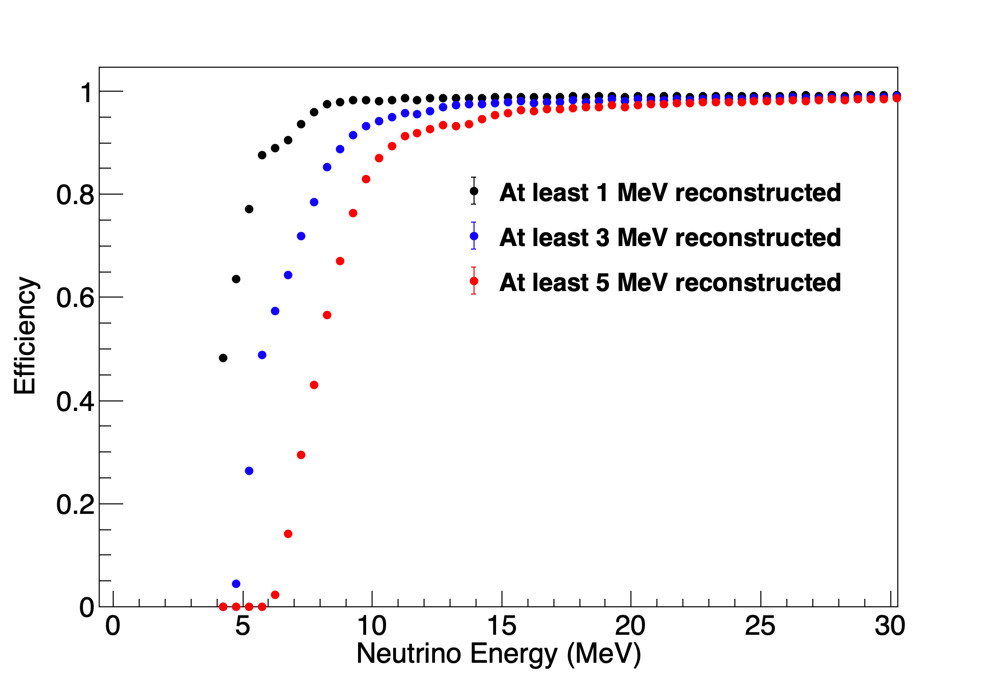}
\includegraphics[width=0.45\textwidth]{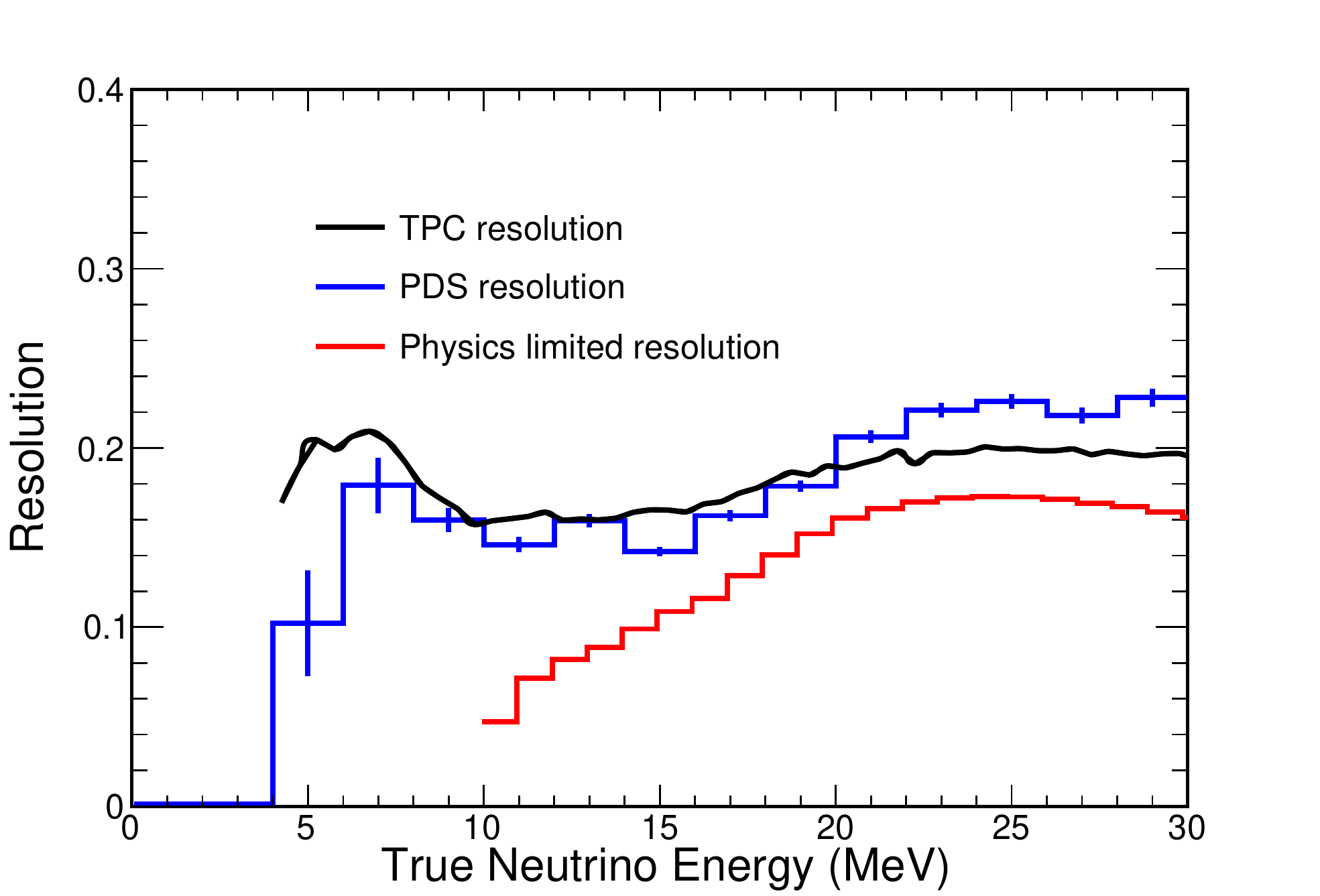}
\caption{Left:
    reconstruction efficiency as a function of neutrino energy for
    MARLEY $\nu_e$CC events, for different minimum required reconstructed
    energy. Right: fractional energy resolution (RMS of the distribution of the fractional difference between reconstructed and true energy with respect to true energy) as a function of
    neutrino energy for TPC tracks corrected for drift attenuation (black) and photon detector
    calorimetry (blue). The red ``physics-limited resolution'' is the ratio of the RMS to the mean of the deposited energy distribution, and assumes
  all energy deposited by final-state particles is reconstructed; the
  finite resolution represents loss of energy from escaping particles (primarily neutrons).  Below 10~MeV the RMS of this distribution is zero. \label{fig:reseff}}

\end{figure*}

Figure~\ref{fig:reseff} shows summarized fractional energy resolution and efficiency performance for
MARLEY events.   Angular resolution performance will be addressed in a
separate publication.

\subsubsection{Backgrounds}

Understanding of cosmogenic~\cite{Zhu:2018rwc} and radiological
backgrounds is also necessary for determination of low-energy event
reconstruction quality and for setting detector
requirements.    The dominant radiological is expected to be $^{39}$Ar, which $\beta$ decays at a rate of $\sim$1~Bq/liter, with an endpoint of $<$1~MeV.
Small single-hit blips from these decays or other
radiologicals may fake de-excitation gammas.  However
preliminary studies
show that these background blips will have a very minor effect on reconstruction of
triggered supernova burst events.
The effects of backgrounds on a data acquisition (DAQ) and
triggering system that satisfies supernova burst triggering
requirements need separate consideration.  These will be the topics of future study.   For studies presented here, the impact of backgrounds on event reconstruction is ignored.

\subsection{Expected Supernova Burst Signal }\label{sec:sn-signals}

\subsubsection{ \snowglobes}\label{snowglobes}

\begin{figure*}
\centering
\includegraphics[width=0.45\textwidth]{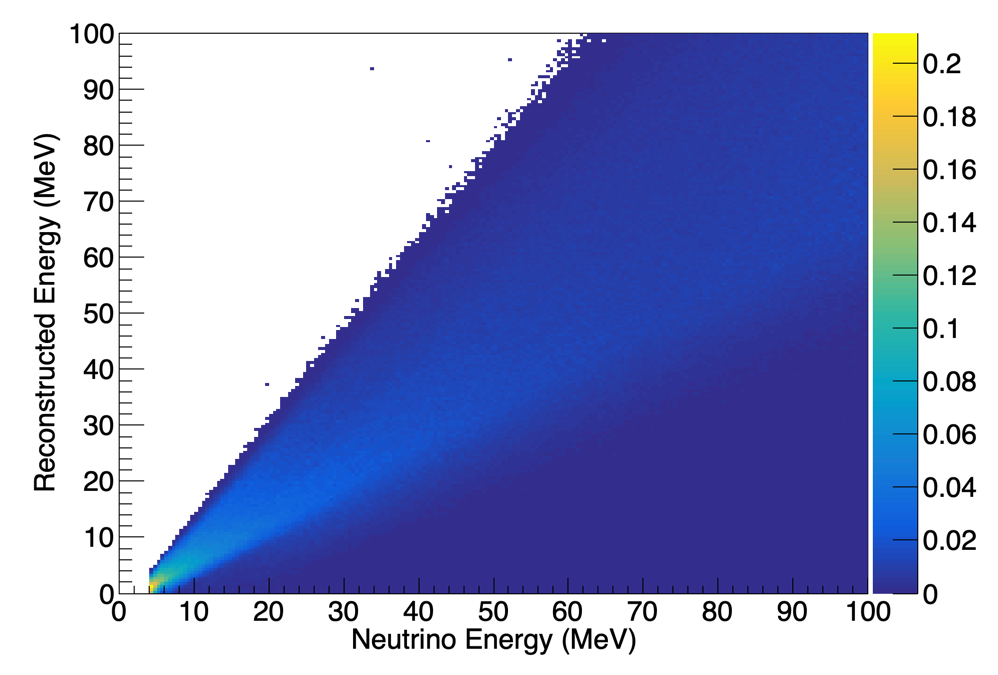}
\includegraphics[width=0.45\textwidth]{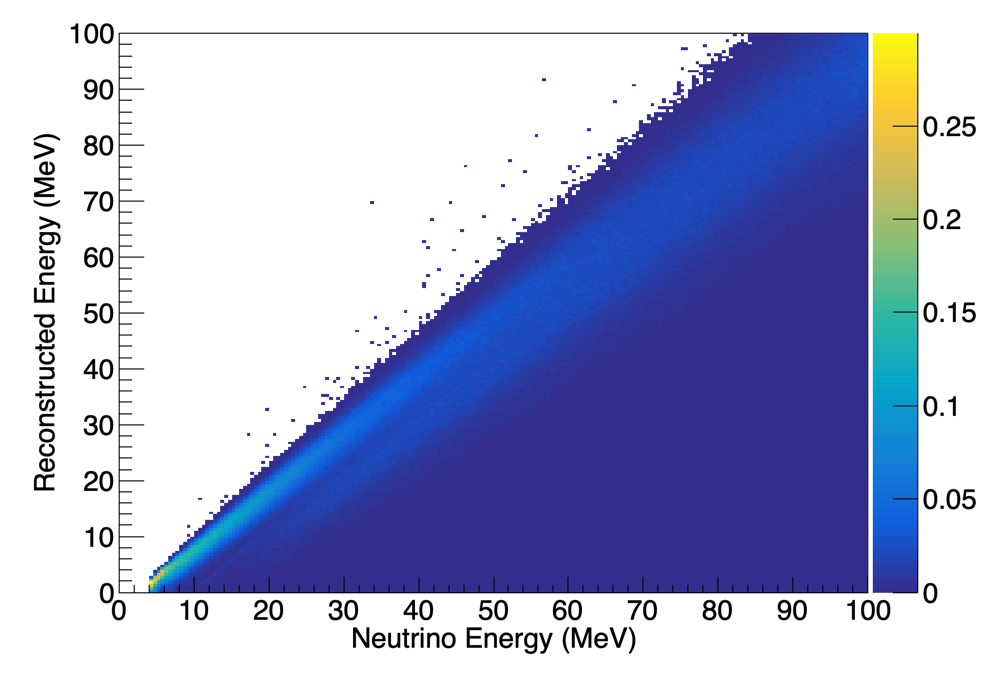}

\caption{Left: transfer matrix for
     \snowglobes~created with monoenergetic $\nu_e$CC MARLEY samples run though
    LArSoft, with the color scale indicating the relative detected charge distribution as a function of
    neutrino energy.  The effects of interaction product distributions
  and detector smearing are both incorporated in this transfer matrix.  The
  right hand plot 
  incorporates an assumed correction for charge attenuation due to
  electron drift in the TPC,  based on Monte Carlo truth position of the
  interaction.   This correction can be made using PDS information.  The drift correction improves energy resolution.\label{fig:marleysmearing}}
\end{figure*}

\begin{figure}[htbp]
\includegraphics[width=0.45\textwidth]{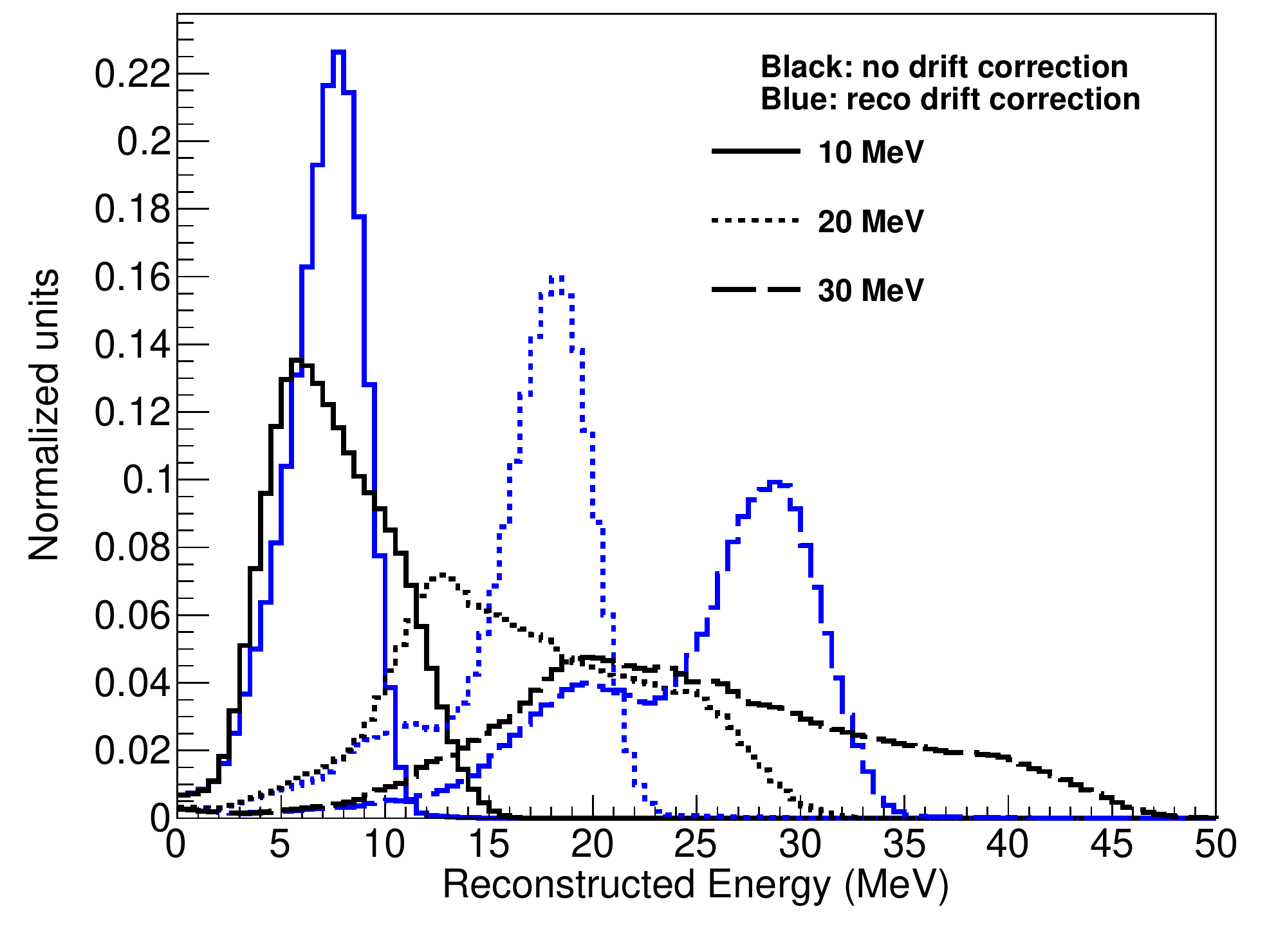}
\caption{\label{fig:transferslices} Observed reconstructed energy distributions for specific interacting neutrino energies (corresponding to columns of the transfer matrices in Fig.~\ref{fig:marleysmearing}), with and without reconstructed photon drift correction.}
\end{figure}

Many supernova neutrino studies done for DUNE so far 
have employed
 \snowglobes~\cite{snowglobes}, a fast event-rate computation tool.  This
uses 
GLoBES front-end software~\cite{Huber:2004ka} to
convolve fluxes with cross sections and detector parameters.  The
output is in the form of both mean interaction rates for each channel as a
function of neutrino energy and mean ``smeared'' rates as a function of
detected energy for each channel (i.e., the spectrum that
actually would be observed in a detector).  
The smearing (transfer) matrices incorporate both
interaction product spectra for a given neutrino energy and detector
response.   Figure~\ref{fig:marleysmearing} shows examples of such  transfer matrices
created
using MARLEY and LArSoft.  They were made by determining the distribution of reconstructed charge using
 a full simulation of the detector response (including the generation,
 transport, and detection of ionization signals and the electronics,
 followed by high-level reconstruction algorithms)
 as a function of neutrino energy in 0.5-MeV steps.
 Each column of a transfer matrix for a given interaction channel
 represents the detector response
 to interactions of monoenergetic neutrinos in the detector.  An electron drift attenuation correction, which can be computed using the reconstructed photon signal (which determines the time of the interaction and hence the drift distance), improves resolution significantly; see Fig.~\ref{fig:transferslices}.
 
Time dependence of a supernova flux in  \snowglobes~can be straightforwardly
handled by providing multiple fluxes divided into different
time bins (see Fig.~\ref{fig:early}), although studies here assume a time-integrated flux.

While \snowglobes~is, and will continue to be, a fast, useful tool,
it has limitations with respect to a full simulation.  One loses correlated
event-by-event angular and energy information, for example; studies of directionality 
require such complete event-by-event information~\cite{ajpointingtalk}.
Nevertheless, transfer matrices generated with full
simulations can be used for fast computation of observed event rates and energy distributions from which useful conclusions can be drawn.

\subsubsection{Expected Event Rates}\label{sec:event_rates}

Table~\ref{tab:argon_events} shows rates calculated  for the dominant interactions in argon for
the ``Livermore'' model~\cite{Totani:1997vj} (included for comparison with literature), the ``GKVM''
model~\cite{Gava:2009pj}, and the ``Garching" electron-capture supernova model~\cite{Huedepohl:2009wh}.\footnote{We are aware that, unlike the model in ~\cite{Huedepohl:2009wh}, the Livermore and GKVM fluxes are not based on full state-of-the-art simulations.  However, they produce results within range of more sophisticated models. Furthermore the fluxes are available in \snowglobes and appear frequently in past literature, so we include them as examples.}  For the first and last, no flavor transitions are assumed
in the supernova or Earth; the GKVM model assumes collective effects in
the supernova.  In general, there is a rather wide variation --- up to
an order of magnitude --- in event rate for different models due to
different numerical treatment (e.g., neutrino transport,
dimensionality), physics input (nuclear equation of state, nuclear
correlation and impact on neutrino opacities, neutrino-nucleus
interactions) and flavor transition effects. In addition, there is intrinsic variation in the nature of the progenitor and collapse mechanism.  Neutrino emission from the supernova may furthermore have an emitted lepton-flavor asymmetry~\cite{Tamborra:2014aua}, so that observed rates may be dependent on the supernova direction.
\begin{table}[h]

\begin{tabular}{c|c|c|c}
Channel & Liver- & GKVM & Garching\\
 & more & & \\ \hline

$\nu_e + ^{40}{\rm Ar} \rightarrow e^- + ^{40}{\rm K^*}$ & 2648  & 3295 & 882 \\ \hline

$\overline{\nu}_e + ^{40}{\rm Ar} \rightarrow e^+ + ^{40}{\rm Cl^*}$ & 224 & 155 & 23 \\ \hline

$\nu_X + e^- \rightarrow \nu_X + e^-$                           & 341 & 206  & 142\\ \hline

Total &  3213 & 3656 & 1047\\ 
\end{tabular}
\caption{Event counts computed with \snowglobes~for different
    supernova models in 40~kton of liquid argon for a core collapse at 10~kpc, for $\nu_e$CC and $\bar{\nu}_e$CC channels and ES ($X$ represents all flavors) on electrons.
    Event rates will simply scale by active detector mass and inverse
    square of supernova distance.   No flavor transitions are assumed for the ``Livermore" and ``Garching" models;  the ``GKVM" model includes collective effects.
    Note that flavor transitions (both standard and collective) will
    potentially have a large, model-dependent effect, as discussed in Sec.~\ref{sec:mh}.\label{tab:argon_events}}
\end{table}

\begin{figure}[htbp]
\includegraphics[width=3in]{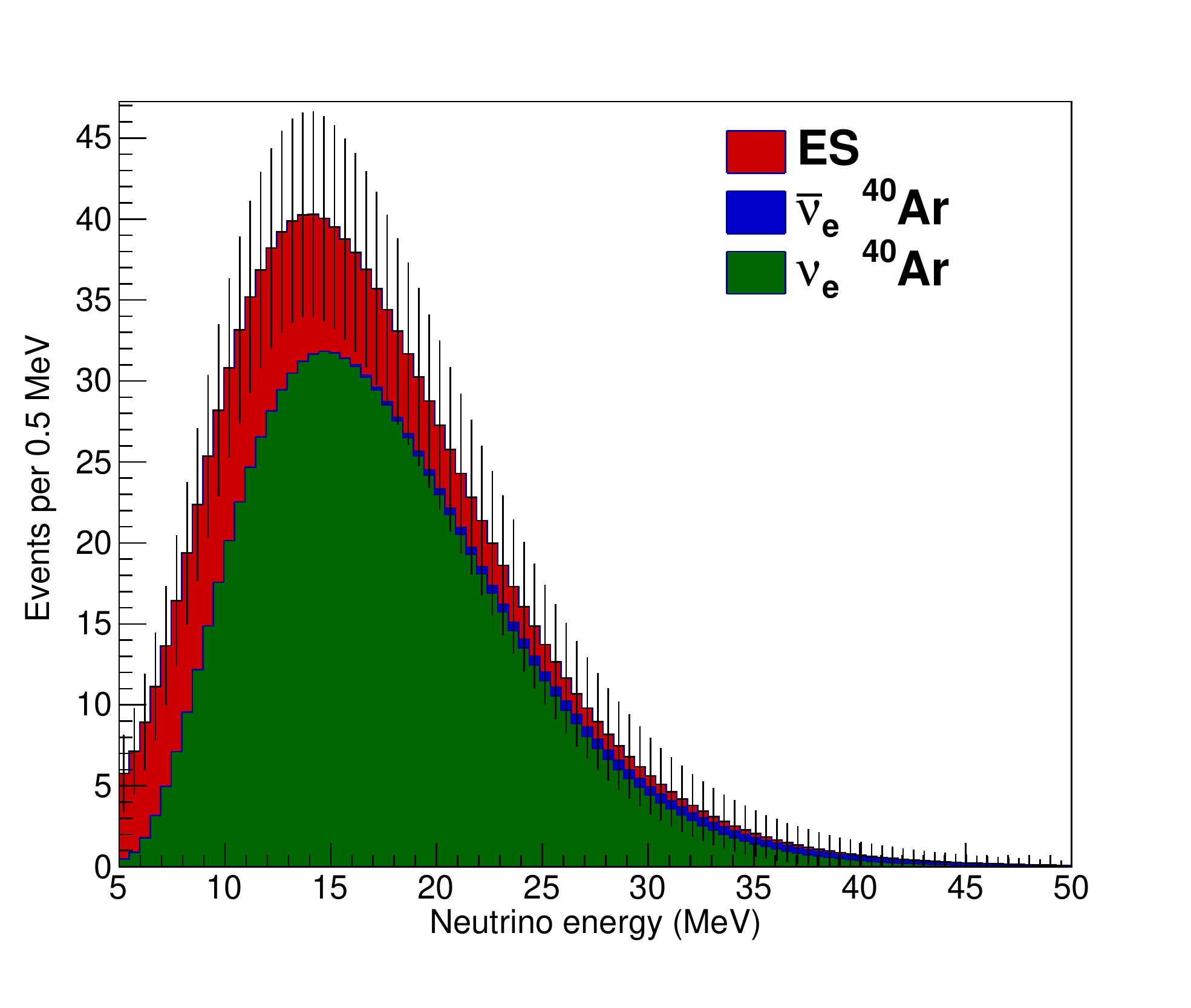}
\includegraphics[width=3in]{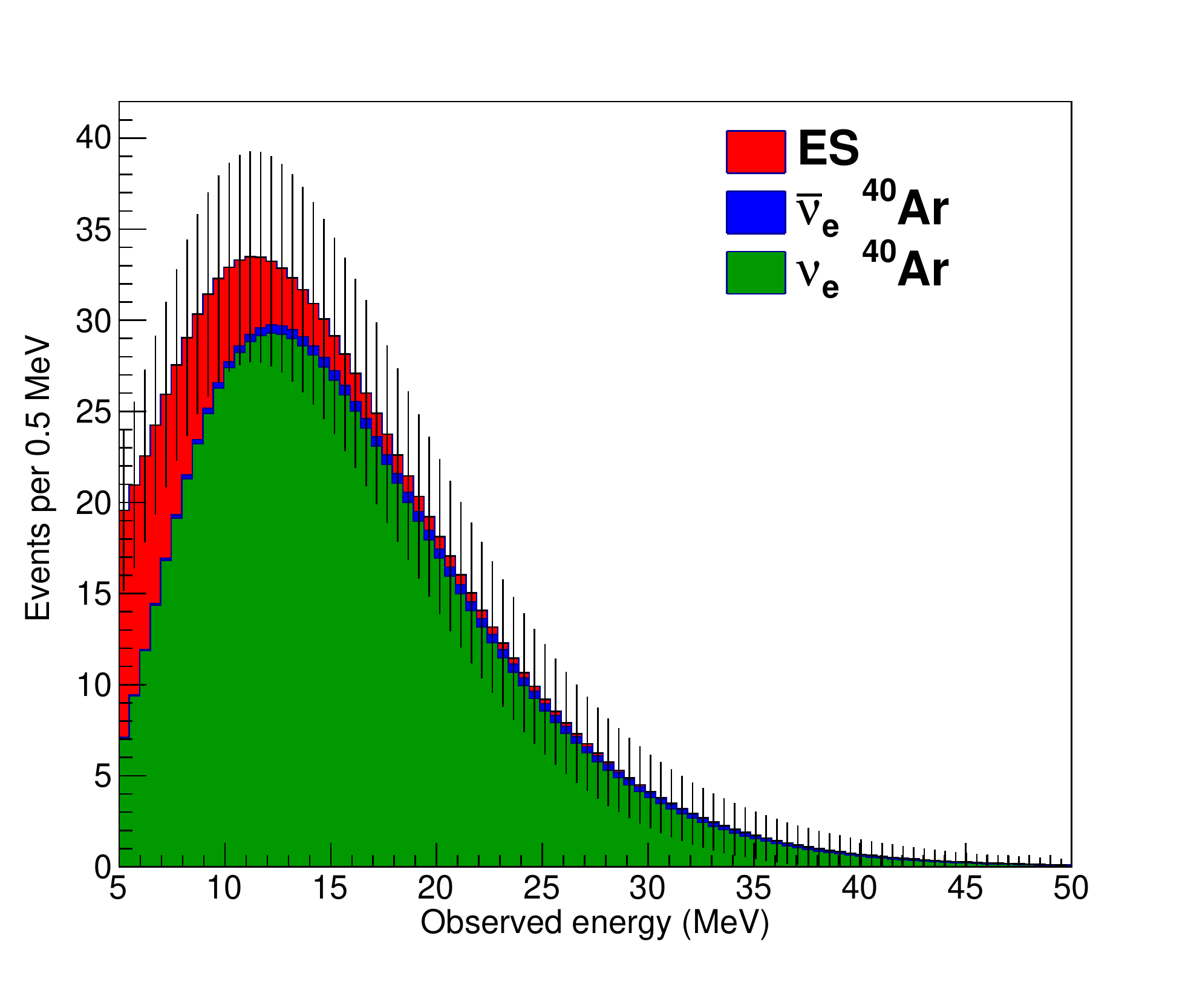}
\caption{Top: Spectrum as a function of interacted neutrino energy computed with~\snowglobes~in 40 kton of liquid argon for the electron-capture supernova~\cite{Huedepohl:2009wh} (``Garching" model) at 10~kpc, integrated over time, and indicating the contributions from different interaction channels.  No oscillations are assumed.  Bottom: expected measured spectrum as a function of observed energy, after detector response smearing. \label{fig:garchingspec}}
\end{figure}

Figure~\ref{fig:garchingspec} shows the expected event spectrum  and the interaction channel breakdown for the ``Garching" model before and after detector response smearing with~\snowglobes.
Clearly, the $\nu_e$
flavor dominates.  Although water and scintillator detectors will
record $\nu_e$ events~\cite{Laha:2013hva,Laha:2014yua}, the $\nu_e$
flavor may not be cleanly separable in these detectors.
Liquid argon is the only future prospect for a large, cleanly tagged supernova $\nu_e$ sample~\cite{Scholberg:2012id}.

\begin{figure*}
\centering
\includegraphics[width=0.8\textwidth]{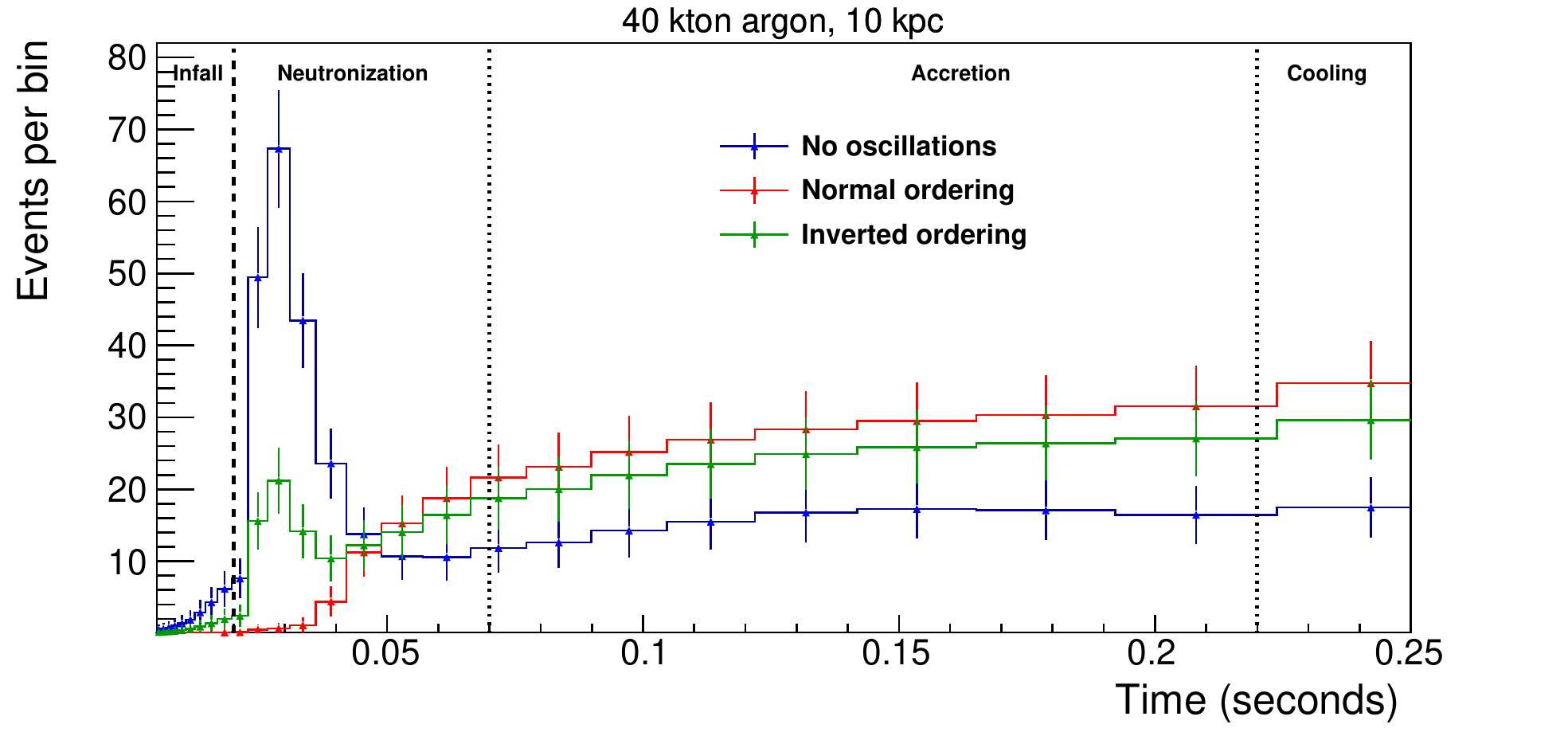}
\caption{Expected event rates as a function of time for the electron-capture model in~\cite{Huedepohl:2009wh} for 40~kton of argon during early stages of the event -- the neutronization burst and early accretion phases, for which self-induced effects are unlikely to be important.  Shown are: the event rate for the unrealistic case of no flavor transitions (blue) and the event rates including the effect of matter transitions for the normal (red)  and inverted (green) orderings.  Error bars are statistical, in unequal time bins.\label{fig:early}}

\end{figure*}

Figure~\ref{fig:early} shows computed event rates
 showing the effect of different mass orderings, using the assumptions in Sec.~\ref{sec:mh}.
MSW-dominated transitions affect
the subsequent rise of the signal over a fraction of a second; 
the time profile will depend on the turn-on of the non-$\nu_e$
flavors.   For this model at 10~kpc there are statistically-significant differences in the time
profile of the signal for the different orderings.

\begin{figure*}
\centering
\includegraphics[width=5in]{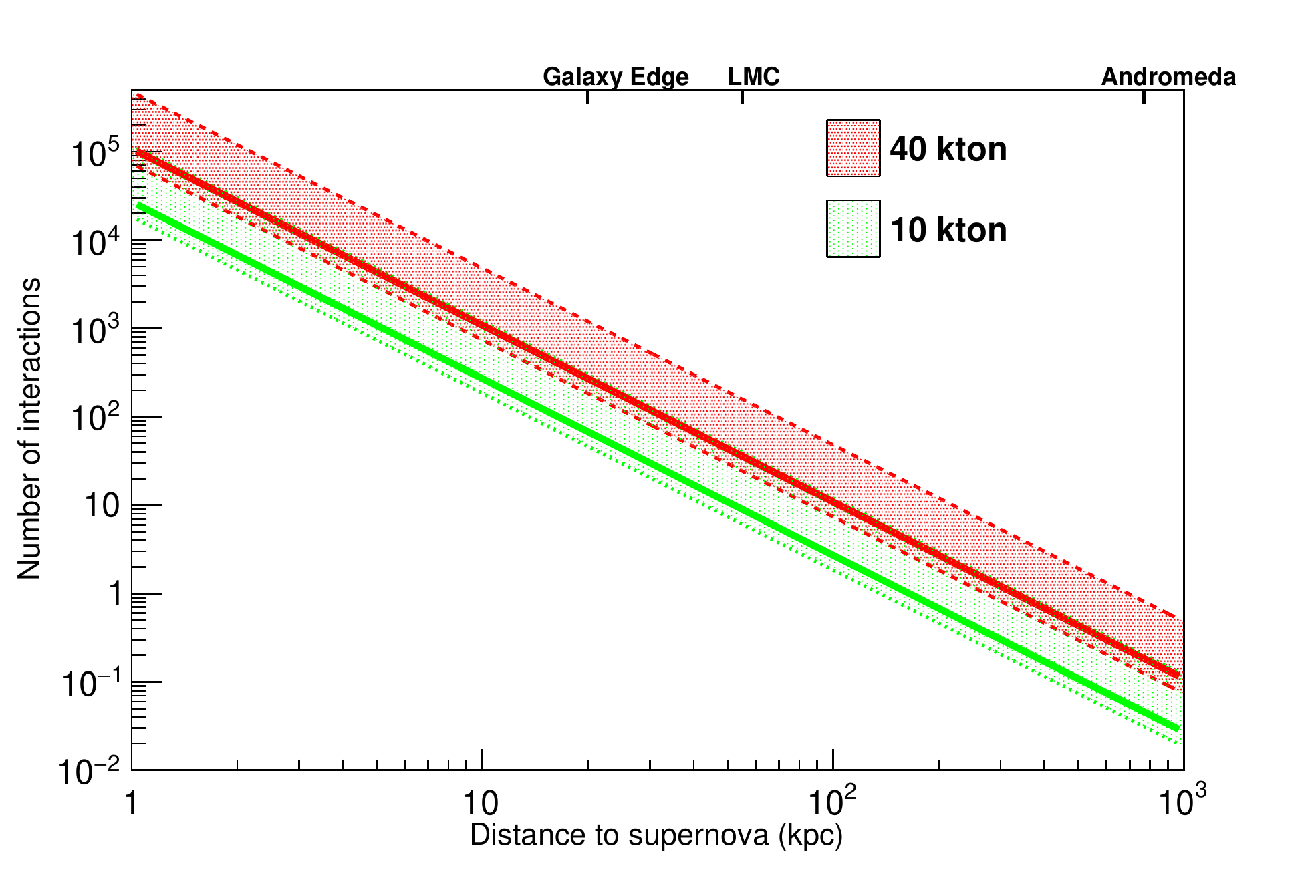}
\caption{Estimated numbers of supernova neutrino
    interactions in DUNE as a function of distance to the supernova,
    for different detector masses ($\nu_e$ events dominate). The red
    dashed lines represent expected events for a 40-kton detector and
    the green dotted lines represent expected events for a 10-kton
    detector. The lines limit a fairly wide range of possibilities for
    pinched-thermal-parameterized supernova flux spectra
    (Equation~\ref{eq:pinched}) with luminosity $0.5\times 10^{52}$
    ergs over ten seconds. The optimistic upper line of a pair gives
    the number of events for average $\nu_e$ energy of $\langle
    E_{\nu_e}\rangle =12$~MeV, and pinching parameter $\alpha=2$;
    the pessimistic lower line of a pair gives the number of events
    for $\langle E_{\nu_e}\rangle=8$~MeV and $\alpha=6$. (Note that
    the luminosity, average energy and pinching parameters will vary
    over the time frame of the burst, and these estimates assume a
    constant spectrum in time. Flavor transitions will also affect the spectra and event rates.) The solid lines represent the integrated number of events for the specific time-dependent neutrino flux model in~\cite{Huedepohl:2009wh} (see Figs.~\ref{fig:params} and \ref{fig:3timescales}; this model has relatively cool spectra and low event rates). Core collapses are expected to occur a few times per century, at a most-likely distance of around 10 to 15 kpc. \label{fig:ratesvsdist}}

\end{figure*}

For a given supernova, the number of signal events scales with detector mass and inverse square of
distance as shown in Fig.~\ref{fig:ratesvsdist}. The standard supernova distance is 10~kpc, which is just beyond the center of the Milky Way. At this distance, DUNE will observe from several hundred to several thousand events.
For a collapse in the
Andromeda galaxy, 780~kpc away, a 40-kton detector would observe a few events at most.

\subsection{Burst Triggering}\label{sec:burst_triggering}

Given the
rarity of a supernova neutrino burst in our galactic neighbourhood and
the importance of its detection, it is essential to develop a
redundant and highly efficient triggering scheme in DUNE. 
In DUNE, the trigger on a supernova neutrino burst can be done using
either TPC or photon detection system information. In both cases, the
trigger scheme exploits the time coincidence of multiple signals over
a timescale matching the supernova luminosity evolution. Development
of such a data acquisition and triggering scheme is a major activity
within DUNE and will be the topic of future dedicated publications.  Both TPC and PD information can be used for triggering, for both SP and DP.  Here are described two concrete examples of preliminary trigger design studies.
Note that the general strategy will be to record data from all channels over a 30-100~second period around every trigger~\cite{Abi:2020loh}, so that the individual event reconstruction efficiency as described in Sec.~\ref{sec:performance} will apply for physics performance.

The first example is a trigger based on the photon detection system of
the DP module.  A real-time algorithm should provide trigger primitives by searching for photomultiplier hits and optical clusters, where the latter combines several hits together based on their time/spatial information. According to simulations, the optimal cluster reconstruction parameters yield a 0.05~Hz radiological background cluster rate for a supernova $\nu_{\text{e}}$CC signal cluster efficiency of 11.8\%. Once the optimal cluster parameters are found, the computation of the supernova neutrino burst trigger efficiency is performed using the minimum cluster multiplicity. This value, set by the radiological background cluster rate and the maximum fake trigger rate (one per month), is $\geq$3 in a 2-second window (time in which about half of the events are expected). Approximately 3/0.118$\simeq$25 interactions must occur in the active volume to obtain approximately 45\% trigger efficiency while maintaining a fake trigger rate of one per month.

\begin{figure}[htbp]
\includegraphics[width=0.98\linewidth]{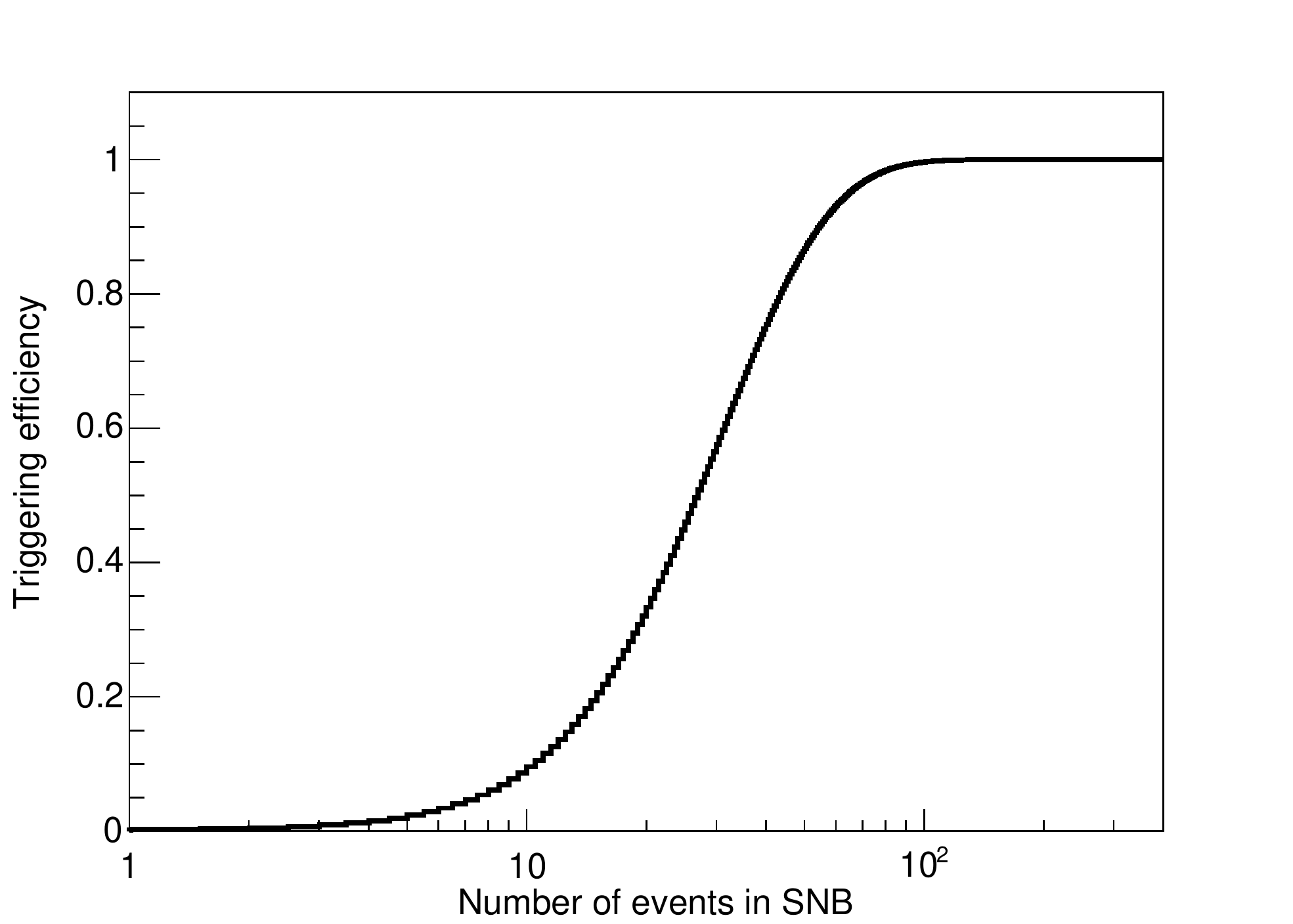}
\caption{Supernova neutrino burst triggering efficiency for the DP photon detectors as a function
  of the number of interactions in one module of the dual phase active volume for the  wavelength-shifting reflective half-foil configuration of the baseline design. \label{fig:triggerdp}}

\end{figure}

The triggering efficiency as a function of the number of supernova neutrino interactions is shown in Fig.~\ref{fig:triggerdp}. 
At 20 kpc, the edge of the Galaxy, about 80 supernova neutrino
interactions in the 12.1-kton active mass (assumed supernova-burst-sensitive mass for a
single DP module) are expected (see Fig.~\ref{fig:ratesvsdist}). Therefore, the DP photon detection system should yield a highly efficient trigger for a supernova neutrino burst occurring anywhere in the Milky Way.

\begin{figure}[htbp]
\includegraphics[width=0.98\linewidth, trim=0 0 0 -30]{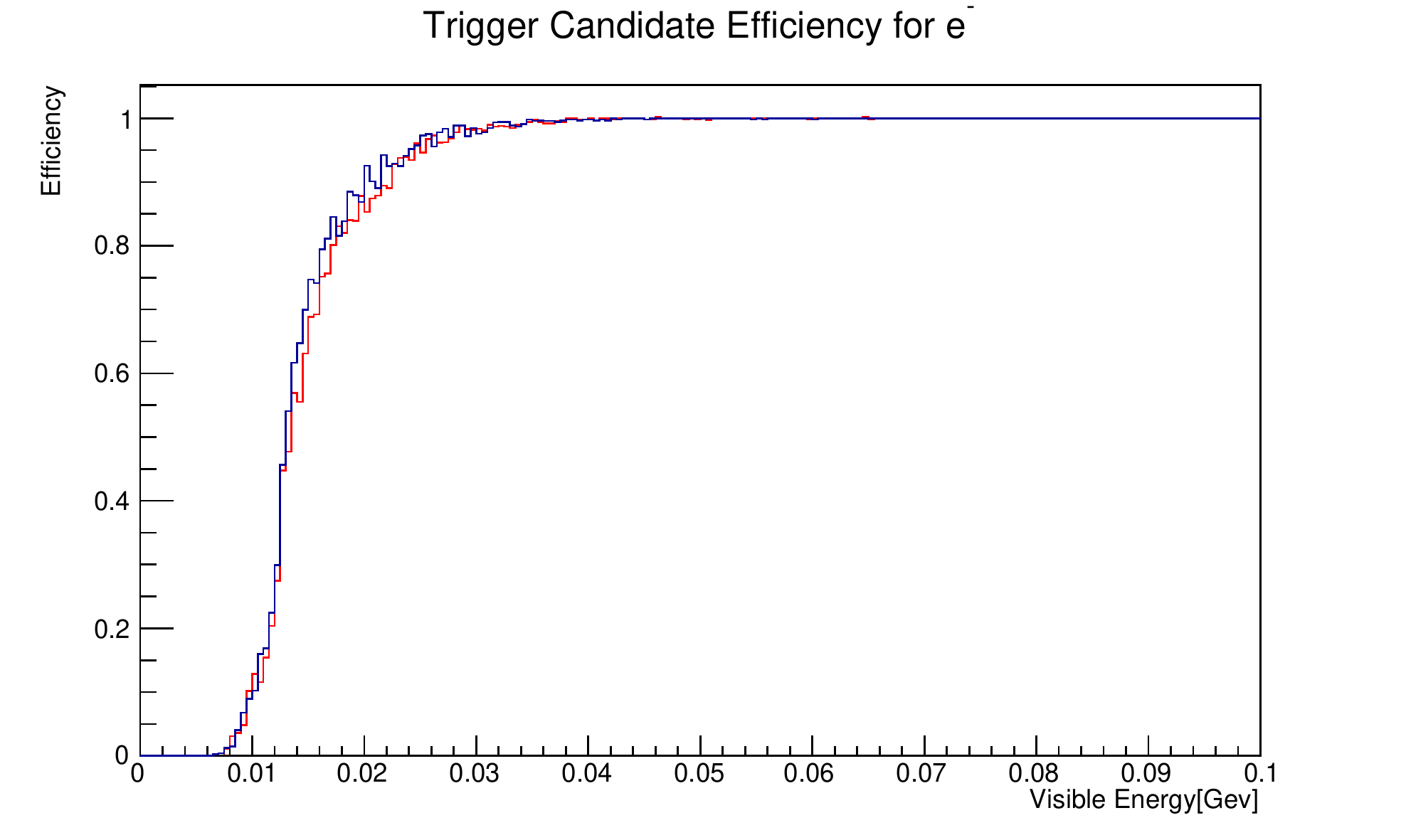}
\includegraphics[width=0.98\linewidth, trim = 0 120 0 120]{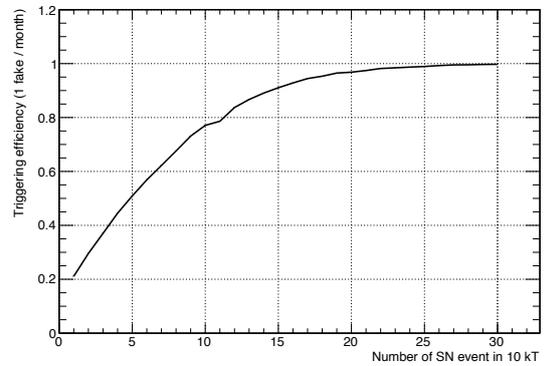}
\vspace{0.4in}
\includegraphics[width=0.4\linewidth, angle=-90, trim=170 80 170 80]{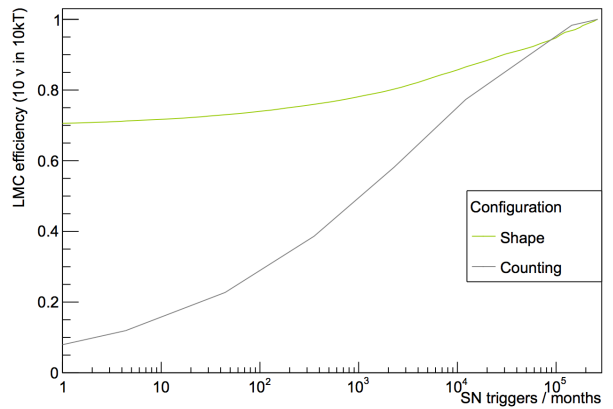}
\vspace{0.4in}
\caption{
Top: Single-interaction efficiency for forming trigger candidates from trigger primitives generated online (in blue) and offline (in red), using SP TPC information, as a function of visible energy for electrons such as those from low-energy $\nu_e$CC scattering on argon.
Middle: Supernova burst trigger efficiency as a function of the number of supernova neutrino interactions
expected in a 10-kton SP module, for a likelihood trigger approach that utilizes sum digitized-charge shape information of trigger
candidates input into the trigger decision. Bottom: Supernova burst trigger efficiency as a function of total (signal and fake) trigger bursts per month, for a supernova burst at the Large Magellanic Cloud, where about 10 neutrino interactions are expected in a 10~kton module (see Fig.~\ref{fig:ratesvsdist}). The efficiency gain with an energy-weighted scheme over a counting-only trigger is significantly improved.\label{fig:tpceffw}}
\end{figure}

The second example considered is a TPC-based supernova neutrino burst
trigger in a SP module (SP photon-based triggering will be considered in a future study).  Such a trigger considering the time coincidence of multiple neutrino interactions over a period of up to 10~seconds yields roughly comparable efficiencies. Figure~\ref{fig:tpceffw} shows efficiencies for supernova bursts obtained in this way for a DUNE SP module and for supernova bursts with an energy and time evolution as shown in Fig.~\ref{fig:params}. Triggering using TPC information is facilitated by a multi-level data selection chain whereby ionization charge deposits are first selected on a per wire basis, using a threshold-based hit finding scheme. This results in low-level trigger primitives (hit summaries) which can be correlated in time and channel space to construct higher-level trigger candidate objects. Low-energy trigger candidates, each consistent with the ionization deposition due to a single supernova neutrino interaction, subsequently serve as input to the supernova burst trigger. Simulations demonstrate that the trigger candidate efficiency for any individual supernova burst neutrino interaction is on the order of 20-30\%; see Fig.~\ref{fig:tpceffw}. However, a multiplicity-based supernova burst trigger that integrates low-energy trigger candidates over $\sim$10~s integration window yields high trigger efficiency out to the galactic edge while keeping fake supernova burst trigger rates due to noise and radiological backgrounds to the required level of one per month or less.  

An energy-weighted multiplicity count scheme further increases efficiency and minimizes fake triggers due to noise and/or radiological backgrounds. This effect is illustrated in Fig.~\ref{fig:tpceffw}, where a nearly 100\% efficiency is possible out to the edge of the galaxy, and 70\% efficiency is possible for a burst at the Large Magellanic Cloud (or for any supernova burst creating $\sim$10 events). This performance is obtained by considering the summed-waveform digitized-charge distribution of trigger candidates over 10 s and comparing to a background-only vs.~background-plus-burst hypothesis. The efficiency gain compared to a simpler, trigger candidate counting-based approach is significant; using only counting information, the efficiency for a supernova burst at the Large Magellanic Cloud is only 6.5\%. These algorithms are being refined to further improve supernova burst trigger efficiency for more distant supernova bursts. Alternative data selection and triggering schemes are also being investigated, involving, e.g., deep neural networks implemented for real-time or online data processing in the DAQ~\cite{bib:docdb11311}.

\subsection{Event Timing in DUNE}
Timing for supernova neutrino events is provided by both the TPC and
the photon detector system.  Basic timing requirements are set by
event vertexing and fiducialization needs. Here we note a few
supernova-specific design considerations.  During the first 50~ms of a
10-kpc-distant supernova, the mean interval between successive
neutrino interactions is $0.5 - 1.7~\rm{ms}$ depending on the model.
The TPC alone provides a time resolution of 0.6~ms (corresponding to the drift time at 500~V/cm),
commensurate with the fundamental statistical limitations at this
distance.  However nearly half of galactic supernova candidates lie
closer to Earth than this, so the rate can be tens or (less likely)
hundreds of times higher.  A resolution of $\mathord{<}1~\mu\rm{s}$,
as already provided by the photon detector system, ensures that DUNE's
measurement of the neutrino burst time profile is always limited by
rate and not detector resolution.  The hypothesized oscillations of
the neutrino flux due to standing accretion shock instabilities would
lead to features with a characteristic time of $\sim$10~ms,
comfortably greater than the time resolution.  The possible neutrino
``trapping notch'' (dip in luminosity due to trapping of neutrinos in
the stellar core)
right before  the start of the neutronization burst has a width of
$1 - 2~\rm{ms}$.  Identifying the trapping notch could be possible for
the closest supernovae (few kpc).

\section{Extraction of Supernova Flux Parameters}\label{sec:flux-params}

This example of a complete study investigates how well it will be possible to fit to the supernova
pinched-thermal flux parameters, to determine, for example, the $\varepsilon$
parameter related to the total binding energy release of the supernova
(proportional to the normalization in Eq.~\ref{eq:pinched}).   Similar
studies in the literature for different detectors
include e.g.,~\cite{GilBotella:2004bv,Minakata:2008nc,Nikrant:2017nya,GalloRosso:2017mdz}.  We examine generically the effect of energy resolution and statistics on the ability to reconstruct flux parameters.

The \snowglobes~package models neutrino signals described by the pinched-thermal
form.  A
forward-fitting algorithm requiring a  \snowglobes-generated energy
spectrum for a supernova at a given distance and a chosen ``true" set of
pinched-thermal parameters $(\alpha^0, \langle E_\nu \rangle^0,
\varepsilon^0)$ was developed. As an example, the true parameter
values are chosen
$(\alpha^0, \langle E_\nu \rangle^0, \varepsilon^0) = (2.5, 9.5,
5\times 10^{52})$, with  $\langle E_\nu \rangle^0$ in MeV and $\varepsilon$ in ergs, assumed
integrated over a ten-second burst.
The study focuses on the $\nu_e$ flux and $\nu_e$CC interactions. The algorithm uses this
spectrum as a ``test spectrum" to compare against a grid of predicted energy
spectra generated with many different combinations of $(\alpha,
\langle E_\nu \rangle, \varepsilon)$. To quantify these comparisons,
the algorithm employs a $\chi^2$ minimization technique to find the
best-fit spectrum.
The $\chi^2$ function is defined as

\begin{equation}
     \chi^2(x) = \sum_{i=1}^{N_b} \frac{\big[N_i(\alpha, \langle E_\nu \rangle, \varepsilon) - N_i(\alpha^0, \langle E_\nu \rangle^0, \varepsilon^0)\big]^2}{\sigma_i^2(x)}
     \label{chi2function}
 \end{equation}
In this expression, $N_b$ is the number of bins for the energy spectra, $N_{i}$ is the number of events in bin $i$, $\sigma_i$ is the uncertainty of the contents in bin $i$ (Poisson statistical uncertainty), $(\alpha, \langle E_\nu \rangle, \varepsilon)$ are the set of model parameters used and $(\alpha^0, \langle E_\nu \rangle^0, \varepsilon^0)$ are the model parameters used to generate the test spectrum.

A test spectrum input into the forward-fitting algorithm produces a
set of $\chi^2$ values for every element in a grid. While the smallest
$\chi^2$ value determines the best fit to the test spectrum, there
exist other grid elements that reasonably fit the test spectrum
according to their $\chi^2$ values. The collection of these grid
elements help determine the expected parameter measurement
uncertainty, represented using sensitivity regions in 2D
flux parameter space.   Three sets of 2D parameter spaces are shown: $(\langle E_\nu \rangle, \alpha)$, $(\langle E_\nu \rangle, \varepsilon)$, and $(\alpha, \varepsilon)$.

\begin{figure*}
\centering
	\includegraphics[scale=0.7]{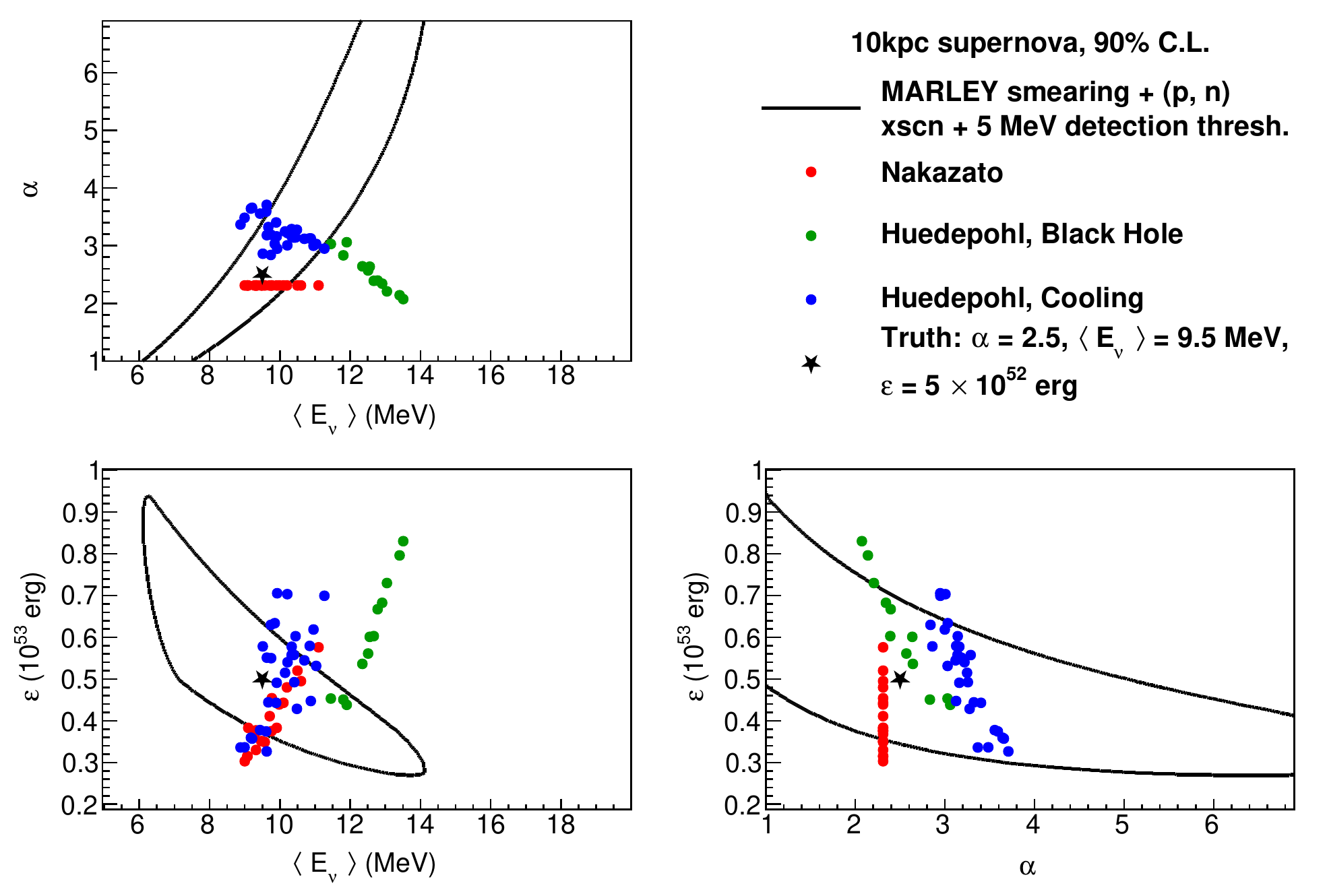}

\caption{Sensitivity regions for the three
    pinched-thermal parameters (90\% C.L.).  The black star represents the assumed true parameters.
  \snowglobes~assumes a cross section
    model from MARLEY, realistic detector smearing from LArSoft, and a step efficiency function with a 5-MeV
    detected energy threshold, for a supernova at 10~kpc. Superimposed
  are parameters corresponding to the time-integrated flux for three different sets of models:
  Nakazato~\cite{Nakazato:2012qf}, Huedepohl black hole formation models, and Huedepohl
  cooling models~\cite{huedepohldb}.  For the Nakazato parameters (for which there is no
  pinching, corresponding to $\alpha=2.3$), the parameters are
  given directly; for the Huedepohl models, they are fit to a
  time-integrated flux.\label{fig:example3params}}

  \end{figure*}

One point in 2D parameter space encompasses several grid elements,
e.g., the $(\langle E_\nu \rangle, \alpha)$ space contains different
$\varepsilon$ values for a given values of $\langle E_\nu \rangle$ and
$\alpha$. To determine the $\chi^2$ value,  
$\varepsilon$ is profiled over to select the grid element with the smallest
$\chi^2$. Sensitivity regions are determined by placing a cut of
$\chi^2 = 4.61$ corresponding to a 90\% coverage probability for two
free parameters.
Figure~\ref{fig:example3params} shows an example of a resulting fit,
where for each set of two parameters, the other is profiled over.  
In this plot, the approximate parameters for three sets of specific models~\cite{Nakazato:2012qf,huedepohldb} are
superimposed, to indicate the expected spread for different assumed progenitor masses,  equations of state, and simulation codes.  A spectral measurement by DUNE would constrain the space of allowed models.

\begin{figure}[htbp]
	\includegraphics[width =0.98\linewidth]{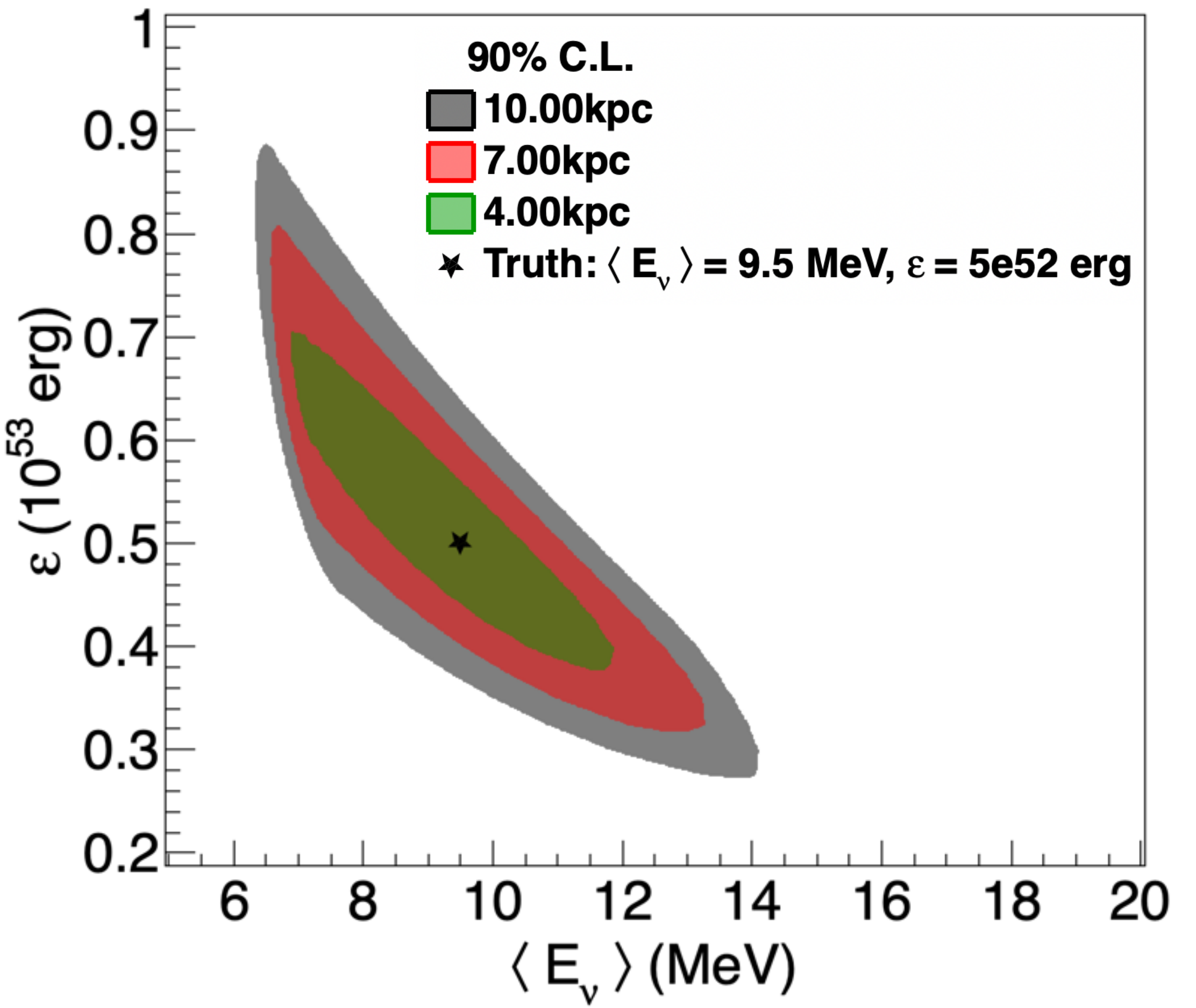}
\caption{Sensitivity regions generated
    in $(\langle E_\nu \rangle, \varepsilon)$ space (profiled over $\alpha$)
    for three different supernova distances (90\% C.L.).  
    \snowglobes~assumes a
   transfer matrix made using MARLEY with a 20\% Gaussian resolution on detected energy, and a step efficiency function with a 5 MeV
    detected energy threshold.\label{fig:exampleAsimovDistance}}

  \end{figure}

\begin{figure}[htbp]
  \includegraphics[width=0.98\linewidth]{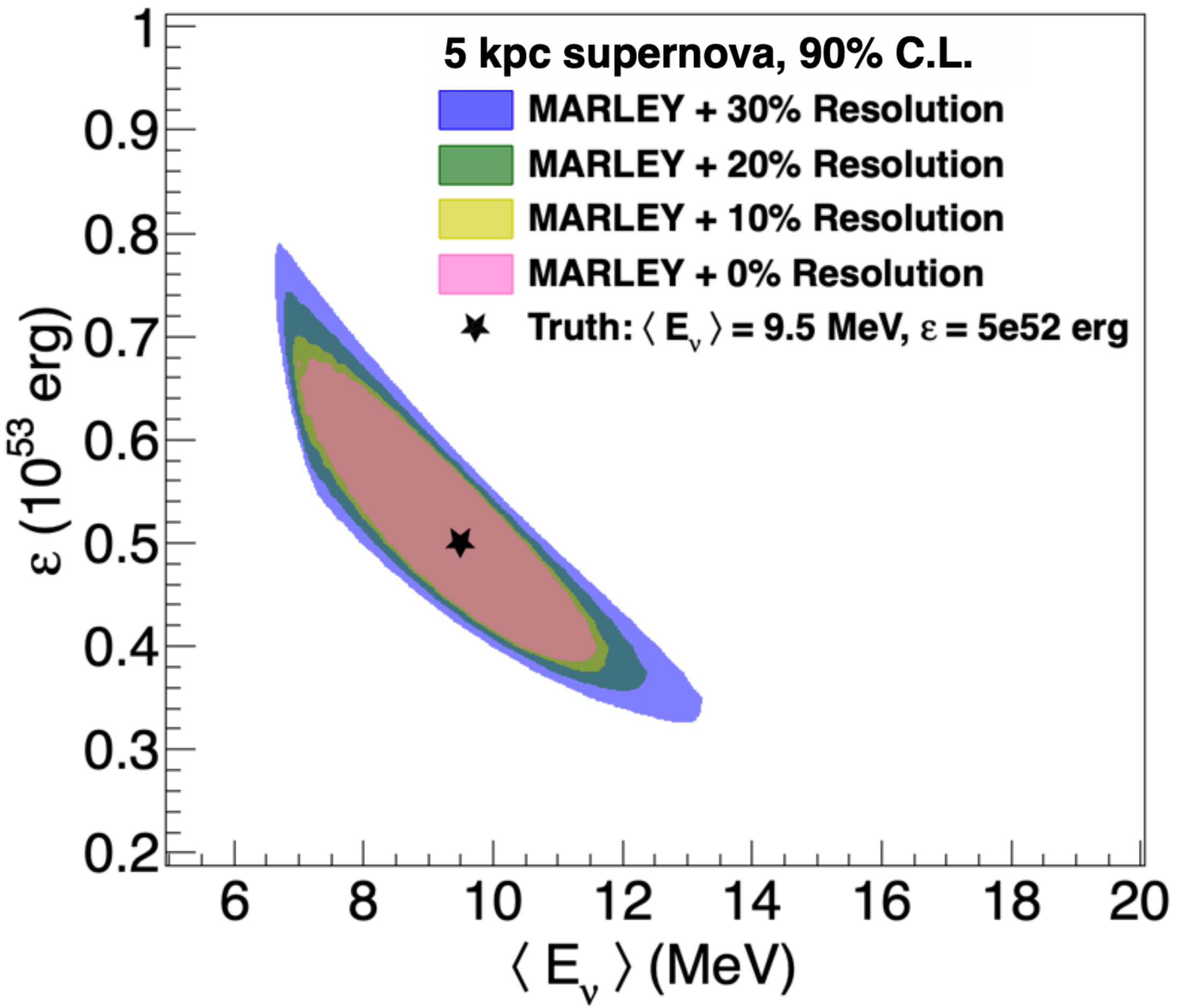}
\caption{90\%~C.L.\ contours for the luminosity and average $\nu_e$ energy
spectral parameters for a supernova
at 5~kpc.  The contours are obtained using the time-integrated
spectrum.  As discussed in the text, the allowed regions
change noticeably but not drastically as one moves from
no detector smearing (pink) to various realistic resolutions (wider regions).\label{fig:specpars}
}

\end{figure}

  Figures~\ref{fig:exampleAsimovDistance} and \ref{fig:specpars} show
  the precision with which DUNE can measure two of the spectral
  parameters, $\varepsilon$, related to the binding energy of the
  neutron star remnant, and $\langle E_{\nu_e}\rangle$, the average
  energy of the $\nu_e$ component, for the time-integrated spectrum
  (profiling over $\alpha$).  Figure~\ref{fig:exampleAsimovDistance}
  shows the statistical effect of different assumed supernova distances
  on determination of the parameters.
  In Fig.~\ref{fig:specpars}, the effect of detector energy resolution
  is examined.  The assumed measured spectrum
  estimated with \snowglobes~takes into account degradation from the
  neutrino interaction process itself ({\em e.g.}, energy lost to
  neutrons), via the MARLEY model.  The colored contours in
  Fig.~\ref{fig:specpars} show increasing levels of assumed detector
  smearing on the measurement of interaction product energy
  deposition.  For 0\% resolution, perfect measurement of the energies
  of interaction products in the detector is assumed.  A 10\% measured
  energy resolution is noticeable but insignificant, and the overall
  precision on the pinched-thermal flux parameters up to 30\% resolution does not
  change dramatically.  According to detector simulation, realistic
  energy resolution is closest to the 20\% level.
  According to Figs.~\ref{fig:exampleAsimovDistance} and~\ref{fig:specpars}, in general, the precision of the measurement of supernova spectral parameters (and the ability to constrain supernova models) is limited more strongly by statistics than by energy resolution.

Given the dominance of $\nu_e$CC events in the
supernova neutrino sample, particle identification is not a
requirement for the primary physics measurements.  However,
additional capability may be possible by identifying separately
NC and ES interactions.

In these studies, we assume that the distance to the core collapse is known.  The interpretation of the $\varepsilon$ parameter as a binding energy will be affected by uncertainty on the distance.  We furthermore assume that mass ordering is known; assumption of incorrect mass ordering results in biases on parameter determination.

\section{Conclusion}

This paper gives an overview of the DUNE experiment's sensitivity to
neutrinos with about 5 MeV up to several tens of MeV, the regime of
relevance for core-collapse supernova burst neutrinos.    This
low-energy regime presents particular challenges for triggering and
reconstruction.  Preliminary DUNE studies show that expected
low-energy background rates should not impede efficient
detection of nearby supernovae.
DUNE's time projection chamber and photon detection systems will both provide information about these
events, and DUNE's software tools have enabled preliminary physics and
astrophysics sensitivity studies.    DUNE will have good sensitivity
to the entire Milky Way and possibly beyond, depending on the
neutrino luminosity of the core-collapse supernova. According to
current understanding, the energy
threshold turn-on is a few MeV deposited energy.  The energy
resolution will be between 10 and 20\% in the few tens of MeV range.  DUNE will be able to measure the supernova $\nu_e$ spectral parameters. By exploiting aspects of a DUNE supernova burst signal, including the time dependence of its energy and flavor profile and non-thermal spectral features, DUNE has the capability to uncover a broad range of supernova and neutrino physics phenomena, including sensitivity to neutrino mass ordering, collective effects, and potentially many other topics.

\section*{Acknowledgements}




%
%
%
%
This document was prepared by the DUNE collaboration using the
resources of the Fermi National Accelerator Laboratory 
(Fermilab), a U.S. Department of Energy, Office of Science, 
HEP User Facility. Fermilab is managed by Fermi Research Alliance, 
LLC (FRA), acting under Contract No. DE-AC02-07CH11359.
%
%
This work was supported by
CNPq, FAPERJ, FAPEG and FAPESP,              Brazil;
CFI, IPP and NSERC,                          Canada;
CERN;
M\v{S}MT,	                                 Czech Republic;
ERDF, H2020-EU and MSCA,                     European Union;
CNRS/IN2P3 and CEA,                          France;
INFN,                                        Italy;
FCT,                                         Portugal;
NRF,                                         South Korea;
CAM, Fundaci\'{o}n ``La Caixa'' and MICINN,  Spain;
SERI and SNSF,                               Switzerland;
T\"UB\.ITAK,                                 Turkey;
The Royal Society and UKRI/STFC,             United Kingdom;
DOE and NSF,                                 United States of America.
%
%
This research used resources of the 
National Energy Research Scientific Computing Center (NERSC), 
a U.S. Department of Energy Office of Science User Facility 
operated under Contract No. DE-AC02-05CH11231.
%

\bibliographystyle{utphys}
\bibliography{tdr-citedb}

\end{document}